\documentclass[longauth]{aa}  
\usepackage{graphicx}
\usepackage{txfonts}
\usepackage{hyperref}
\usepackage{gensymb}
\usepackage{float}

\newcommand{\ngist}{\textsc{nGIST}}
\newcommand{\gist}{\textsc{GIST}}
\newcommand{\ppxf}{\textsc{pPXF~}}
\newcommand{\update}[1]{#1}
\newcommand{\updatetwo}[1]{#1}
\begin{document}

   \title{The GECKOS survey: Jeans anisotropic models of edge-on discs uncover the impact of dust and kinematic structures}

   \author{T. H. Rutherford\inst{1,2},
          A. Fraser-McKelvie\inst{1},
          E. Emsellem\inst{1},
          J. van de Sande\inst{3},
          S. M. Croom\inst{2},
          A. Poci\inst{4},
          M. Martig\inst{5},
          D. A. Gadotti\inst{6},
          F. Pinna\inst{7,8},
          L. M. Valenzuela\inst{9},
          G. van de Ven\inst{10},
          J. Bland-Hawthorn\inst{2},
          P. Das\inst{11},
          T. A. Davis\inst{12},
          R. Elliott\inst{13},
          D. B. Fisher\inst{13},
          M. R. Hayden\inst{14},
          A. Mailvaganam\inst{15,16},
          S. Sharma\inst{17},
          T. Zafar\inst{15}
          }

   \institute{European Southern Observatory, Karl-Schwarzschild-Stra\ss e 2,
              Garching, 85748\\
              \email{trut2989@uni.sydney.edu.au}
              \and
              Sydney Institute for Astronomy, School of Physics, A28, The University of Sydney, NSW, 2006, Australia
              \and
              School of Physics, University of New South Wales, NSW, 2052, Australia
              \and
              Sub-department of Astrophysics, Department of Physics, University of Oxford, Denys Wilkinson Building, Keble Road, Oxford OX1 3RH
               \and
              Astrophysics Research Institute, Liverpool John Moores University, 146 Brownlow Hill, Liverpool L3 5RF, United Kingdom
              \and
              Centre for Extragalactic Astronomy, Department of Physics, Durham University, South Road, Durham DH1 3LE, United Kingdom
              \and
              Instituto de Astrof\'isica de Canarias, Calle V\'ia L\'actea s/n, E-38205 La Laguna, Tenerife, Spain
              \and
              Departamento de Astrof\'isica, Universidad de La Laguna, Av. del Astrof\'isico Francisco S\'anchez s/n, E-38206, La Laguna, Tenerife, Spain
              \and
              Universit\"ats-Sternwarte, Fakult\"at für Physik, Ludwig-Maximilians-Universit\"at M\"unchen, Scheinerstr. 1, 81679 M\"unchen, Germany
              \and
              Department of Astrophysics, University of Vienna, T\"urkenschanzstra\ss e 17, 1180 Vienna, Austria
              \and 
              Astrophysics Research Group, University of Surrey, Guildford, Surrey, GU2 7XH, United Kingdom
              \and
              Cardiff Hub for Astrophysics Research \& Technology, School of Physics \& Astronomy, Cardiff University, Queens Buildings, Cardiff CF24 3AA, United Kingdom
              \and
              Centre for Astrophysics and Supercomputing, Swinburne University of Technology, PO Box 218, Hawthorn, VIC 3122, Australia
              \and
              Homer L. Dodge Department of Physics \& Astronomy, University of Oklahoma, 440 W. Brooks St, Norman, OK 73019, USA
              \and
              School of Mathematical and Physical Sciences, Macquarie University, NSW 2109, Australia
              \and
              Macquarie University Astrophysics and Space Technologies Research Centre, Sydney, NSW 2109, Australia
              \and
              Space Telescope Science Institute, 3700 San Martin Drive, Baltimore, MD 21218, USA
             }

   \date{}
\titlerunning{GECKOS Kinematic Structure and JAM Models}
\authorrunning{Rutherford et al.}
 
  \abstract  
   {The central regions of disc galaxies host a rich variety of stellar structures: nuclear discs, bars, bulges, and boxy-peanut (BP) bulges. These components are often difficult to disentangle, both photometrically and kinematically, particularly in star-forming galaxies where dust obscuration and complex stellar motions complicate interpretation. In this work, we use data from the GECKOS-MUSE survey to investigate the impact of dust on axisymmetric Jeans Anisotropic \updatetwo{Multi-Gaussian Expansion} (JAM) models, and assess their ability to recover kinematic structure in edge-on disc galaxies. We construct JAM models for a sample of seven edge-on ($i \gtrapprox 85^\circ$) galaxies that span a range of star formation rates, dust content, and kinematic complexity. We find that when dust is appropriately masked, the disc regions of each galaxy are fit to $\chi^2_{\text{reduced}}\leq 5$. We analyse two-dimensional residual velocity fields to identify signatures of non-axisymmetric structure. We find that derived dynamical masses are constant within 10\% for each galaxy across all dust masking levels. In NGC\,3957, a barred boxy galaxy in our sample, we identify velocity residuals that persist even under aggressive dust masking, aligned with bar orbits and supported by photometric bar signatures. We extend this analysis to reveal a bar in IC\,1711 and a possible side-on bar in NGC\,0522. Our results highlight both the capabilities and limitations of JAM in dusty, edge-on systems and attempt to link residual velocities to known non-axisymmetric kinematic structure.}

   \maketitle
%

\section{Introduction}

The evolution of massive disc galaxies in the local Universe from their formation until today is a picture that includes significant complexity. The $\Lambda$CDM paradigm asserts that gas collapse within a dark matter halo and hierarchical structure formation are how galaxies build up their mass \citep{1978MNRAS.183..341W,2002NewA....7..155S,2003ApJ...597...21A,2010ApJ...724..915H}, with simulations showing that galaxies experience minor and major mergers over their lifetimes \citep[e.g.][]{2009ApJ...699L.178N,2009A&A...501L...9D,2017MNRAS.464.3850L,2023MNRAS.522....1N}. Minor mergers \citep[e.g.][]{2017ApJ...837...68C,2020MNRAS.493.3778S}, major mergers \citep[e.g.][]{2022MNRAS.509.4372L,2024MNRAS.529..810R}, gas accretion \citep[e.g.][]{2007ApJ...668...94H,2008MNRAS.386..935F,2019ApJ...875...54H,2024A&A...687A.115B}, stellar feedback \citep[e.g.][]{2014MNRAS.443.2092U,2022A&A...666A.161B}, AGN feedback \citep[e.g.][]{2020MNRAS.491.4462D,2024NatAs...8.1443D} and environmental processes \citep[e.g.][]{2011MNRAS.416.1680C,2021ApJ...918...84R} are additionally capable of transforming the stellar motions of stars within a galaxy, primarily in the growth of dispersion-supported bulge-like structures, along with scattering into thick-disc orbits and halo orbits (e.g. MW) \citep[e.g.][]{2011IAUS..271..160B,2012MNRAS.423.1544S,2013MNRAS.433.2986W,2020MNRAS.493.3778S,2022MNRAS.516.3569B}. With the advent of integral-field spectroscopy (IFS) (e.g. SAURON, \citealt{2001MNRAS.326...23B}, MUSE, \citealt{2010SPIE.7735E..08B}, SAMI, \citealt{2012MNRAS.421..872C}), it is now feasible to examine the spatially resolved stellar kinematic maps of massive disc galaxies and link them to other physical properties and merger history.

Photometric studies have revealed the existence of many morphological substructures in galaxies. The largest diversity (and greatest potential for photometric superposition) is found in the central regions of discs, as a result of bars, nuclear discs, and boxy-peanut bulges \citep[e.g.][]{2005ApJ...632..217S,2011ApJ...743L..13C,2015MNRAS.454.3843A,2017A&A...604A..30N,2019ApJ...872....5S,2019MNRAS.489.4992D}. Stellar bars can form either in the centre of spiral galaxies in isolation \citep[e.g.][]{1971ApJ...168..343H,1973ApJ...186..467O}, or be induced through tidal forces arising from galaxy-galaxy interactions and mergers \citep[e.g.][]{1987MNRAS.228..635N,1991A&A...243..118S}. Bars can further funnel gas to the centre of galaxies, forming nuclear discs \citep[e.g.][]{1991MNRAS.252..210B,1992MNRAS.259..345A,1995ApJ...449..508P,2006MNRAS.369..529F,2015A&A...575A...7W,2015ApJ...804..139D,2019MNRAS.482L.118M,2020A&A...643A..14G,2020A&A...643A..65B,2024A&A...687A..53V}, and evolve to buckle vertically out of the disc plane \citep[e.g.][]{1981A&A....96..164C,1999AJ....118..126B,2017MNRAS.468.2058E,2019MNRAS.490.4721K}, giving rise to boxy-peanut (BP) bulges. Evidently, the study of bulges, bars, nuclear discs, and boxy-peanut bulges can assist in tracing the evolutionary history of a galaxy. However, while each component manifests itself photometrically, the integrated two-dimensional projection of light we observe on the sky represents them all combined, creating a very degenerate problem. Although some work has been carried out using major axis surface brightness profiles to identify bars at certain position angles \citep[e.g.][]{1970ApJ...160..811F,2000A&AS..145..405L,2000A&A...362..435L,2006MNRAS.370..753B}, one key way to disentangling these components lies in the stellar kinematics, where a line-of-sight velocity distribution can be determined for each spaxel of a galaxy.

Building on previous work \citep[e.g.][]{1992MNRAS.259..345A,2005ApJ...626..159B,2015MNRAS.450.2514I,2018ApJ...854...65L}, \cite{2025A&A...700A.237F} employed a Gauss-Hermite parametrisation of stellar kinematics ($V$, $\sigma$, $h_3$, $h_4$) \citep{1993ApJ...407..525V,1993MNRAS.265..213G} to qualitatively classify central kinematic structure (e.g. bars, nuclear discs) in 12 edge-on disc galaxies from the GECKOS \footnote{Generalising Edge-on galaxies and their Chemical bimodalities, Kinematics, and Outflows out to Solar environments} \citep{2024IAUS..377...27V} sample. However, the physical interpretation of stellar kinematics can be challenging. While Gauss-Hermite parameters quantify the shape of the line of sight velocity distribution (LOSVD), and are sensitive to underlying kinematic structures \citep[e.g.][]{2004AJ....127.3192C,2005ApJ...626..159B,2020MNRAS.494.5936F}, these quantities are shaped by a combination of intrinsic galaxy properties, such as the mass distribution and velocity anisotropy, and projection effects, such as inclination and line-of-sight integration. This is where dynamical modelling becomes essential, as it can turn observed velocity maps into physically interpretable quantities.

In particular, axisymmetric Jeans Anisotropic Models \citep[JAM,][]{2008MNRAS.390...71C,2020MNRAS.494.4819C} have been widely applied to early-type and passive disc galaxies \citep[e.g.][]{2013MNRAS.432.1709C,2017ApJ...838...77L,2019ApJ...878...57E}, which are typically well suited to axisymmetric modelling due to their lack of strong internal kinematic structure (e.g. bars) and low levels of dust obscuration. To separate the velocities of JAM models into ordered and random motions, it is necessary to make assumptions about the velocity anisotropy. However, enclosed mass profiles and stellar mass-to-light ratios can be recovered from just the $V_{\text{rms}}$ field, and JAM models can still reliably recover these for galaxies with modest velocity non-axisymmetries \citep[e.g.][]{2012MNRAS.424.1495L,2016MNRAS.455.3680L}.

A complete picture of galaxy evolution requires kinematic models of not just passive galaxies, but a full sample across all star formation rates. In particular, the GECKOS \citep[][van de Sande et al. in prep]{2024IAUS..377...27V} sample, provides a particularly difficult opportunity, as they were selected with a $>$2 dex range in star formation rates. In this context, JAM typically struggles with strongly star-forming galaxies, particularly those with complex kinematic structures \citep[e.g.][]{2017MNRAS.464.4789M}. Dust further complicates modelling, particularly in the edge-on case, where it attenuates light from the far side of the disc, preferentially obscures dynamically cold components, and effectively alters the stellar populations probed. Moreover, the stellar orbits of bars and multiple kinematically-decoupled discs (e.g. nuclear discs) are complex and non-axisymmetric \citep[e.g.][]{2002MNRAS.333..847S,2002MNRAS.333..861S,2008gady.book.....B,2016ApJ...818..141V,2021A&A...648L...4T}. The $x_1$ orbits in stellar bars \citep{1980A&A....92...33C}, for example, generate excess non-axisymmetric line-of-sight velocities at the bar ends \citep[e.g.][]{1992MNRAS.259..345A,1993RPPh...56..173S,2017A&A...606A..47F,2024ApJ...976..220K} that are fundamentally inconsistent with an axisymmetric model. Despite these challenges, modelling GECKOS galaxies comprehensively across a range of star-formation rates and dust obscuration levels provides an opportunity to understand how dust impacts derived dynamical parameters for edge-on discs, and further allows us to probe how the combination of structures such as bars and dynamically cold discs impact kinematic measurements.

Orbit-superposition techniques such as Schwarzschild modelling \citep{1979ApJ...232..236S,2020ascl.soft11007J} offer a more general approach, capable of handling \update{orbits} arising from within a non-axisymmetric potential \citep[e.g.][]{2005MNRAS.357.1113K,2008MNRAS.385..647V,2013MNRAS.434.3174V,2015MNRAS.452....2K,2020ApJ...889...39V,2020ascl.soft11007J,2022ApJ...941..109T}, and have recently shown success in reproducing barred and multi-component structures \citep{2024MNRAS.534..861T}. Other techniques such as asymmetric drift correction \citep{2018MNRAS.477..254L} have also been applied, however, these techniques are very computationally demanding, and suffer from the same dust obscuration issues as JAM.

To improve our understanding of how dust and kinematic structure affects dynamical modelling of edge-on galaxies, we have two goals in this paper: 1) to examine and quantify the effect of dust on the goodness-of-fit criterion and returned dynamical parameters of JAM models applied to edge-on discs, and 2) to explore the diversity of kinematic structure at the centre of GECKOS galaxies. To this end, we propose a simple experiment: what happens if we attempt to model an edge-on galaxy with only simple, axisymmetric disc components? The outer regions of unflared disc galaxies are well modelled (i.e. stellar mass and circular velocity can be well recovered) in most cases with a dynamically cold "thin" disc, with or without the addition of a dynamically warmer "thick" disc \citep{2017MNRAS.464.1903K,2018MNRAS.477..254L}. We wish to understand how the velocity residuals obtained after subtracting a JAM model from the data are affected by dust, and if we can link any coherent residual structure to non-axisymmetric kinematic components present. Edge-on galaxies are the perfect test bed for such an experiment, as the greatest component of line-of-sight (LoS) velocity is available to us, dust effects are maximised, and central structure light (i.e. structures that `bulge' out of the disc) should not be superimposed with disc light.

In this paper, we construct axisymmetric JAM models of seven galaxies from the GECKOS survey, representative of a diversity of star formation rates, dust levels and kinematic structures. In Section \ref{sec:data}, we describe the GECKOS data, our sample selection, and surface brightness modelling. In Section \ref{sec:methods} we briefly describe the JAM method and how we applied it to our work. In Section \ref{sec:results} we describe how different dust masks affect the goodness of fit of our models, and describe coherent structures visible in the velocity residuals of our least dusty galaxy, NGC\,3957. In Section \ref{sec:discussion} we quantify the impact of dust on the dynamical parameters returned by JAM models, and extend the analysis of kinematic structures to the remaining galaxies in our sample. Throughout this paper, we use $\Lambda$CDM cosmology, with $\Omega_m$ = 0.30, $\Omega_\Lambda$ = 0.70, and $H_0$ = 70 km s$^{-1}$ Mpc$^{-1}$, and a \cite{2003PASP..115..763C} stellar initial mass function.
\section{Data}
\label{sec:data}
The GECKOS survey is a European Organisation for Astronomical Research in the Southern Hemisphere (ESO) Very Large Telescope (VLT)/Multi-Unit Spectroscopic Explorer (MUSE) large programme. GECKOS targets 36 edge-on disc galaxies, aiming for a signal-to-noise ratio (S/N) of 40 \AA$^{-1}$ per Voronoi bin \citep{2003MNRAS.342..345C}, extending to a surface brightness isophote of $\mu_g = 23.5$ mag arcsec$^{-2}$ - comparable to the Sun’s location in the Galactic disc \citep{2007A&A...462..965M}. Targets were selected within a heliocentric distance range of 15 to 70 Mpc from the S4G survey \citep[25/36;][]{2010PASP..122.1397S} and HyperLeda \citep[11/36;][]{2014A&A...570A..13M}. Eight galaxies in the sample possess archival data, and we extend on these to reach the required surface brightness constraints.

\subsection{Sample selection}
\label{sec:data}
\begin{figure}
    \centering
    \includegraphics[width=\linewidth]{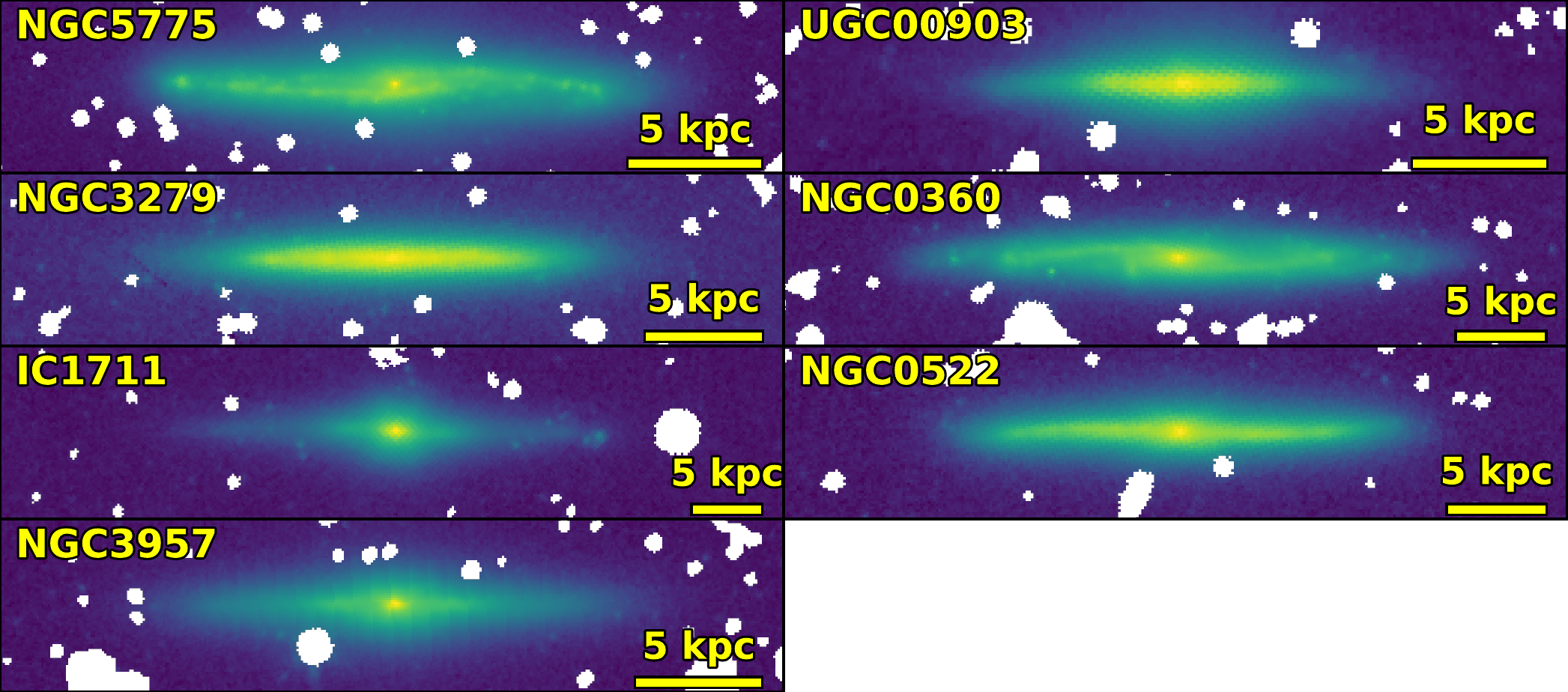}
    \caption{The Spitzer IRAC 3.6$\mu$m imaging of our sample of seven GECKOS galaxies. The images are scaled with an inverse hyperbolic sine (arcsinh) scaling, and masked regions (bright stars, imaging artefacts) are shown in white. There are clear star-forming spiral arms in NGC\,5775 and NGC\,0360.}
    \label{fig:sample_imaging}
\end{figure}

We selected galaxies for this study based on two criteria: (i) they were observed and reduced by December 2023, and (ii) archival Spitzer IRAC 3.6$\mu$m photometry is available for use in Multi-Gaussian Expansion (MGE) modelling (see Section \ref{sec:mge}). We attempt to avoid dust attenuation in our mass and luminosity models by deriving them from Spitzer IRAC 3.6$\mu$m imaging. These constraints result in a pilot sample of seven galaxies: NGC\,5775, UGC\,00903, NGC\,3279, NGC\,0360, IC\,1711, NGC\,0522, NGC\,3957. NGC\,5775 and UGC\,00903 exhibit star formation rates exceeding that of the Milky Way, NGC\,3279 and NGC\,0360 have rates comparable to the Milky Way, while IC\,1711, NGC\,0522, and NGC\,3957 show lower star formation activity. The Spitzer IRAC $3.6\mu$m imaging is presented in Figure \ref{fig:sample_imaging}. Although this sample is limited in size, the sample covers a diversity of star formation rates, dust content, and kinematic structures. The properties of this sample are summarised in Table \ref{tab:sample_properties}, and provide a useful test-bed for exploring the impact of dust and non-axisymmetric structures on dynamical modelling. We note here that the MUSE spatial coverage varies from galaxy to galaxy, impacting the distribution of \update{E(B–V)} values observed in each system.

\begin{table*}
\caption{\updatetwo{Sample Properties.}}    
\label{tab:sample_properties}      
\centering                          
\begin{tabular}{l l p{2.5cm} p{3.5cm} p{3cm}}       
\hline\hline                
ID & SFR [$M_\odot$/yr] & Median \update{E(B–V)} & Kinematic Structures & Distance [Mpc]\\    
\hline
   NGC\,5775 & 6.84 &  0.293 & - & 18.9\\
   UGC\,00903 & 4.17 & 0.207 & Counter-rotating disc & 37.7\\
   NGC\,3279 & 1.99 & 0.117 & Close to pure disc & 29.9\\
   NGC\,0360 & 1.46 & 0.189 & - & 31.2\\
   IC\,1711 & 1.08 & 0.107  & Boxy-peanut bulge and nuclear disc & 44.9\\
   NGC\,0522 & 0.64 & 0.076 &  bulge & 36.2\\
   NGC\,3957 & 0.27 &  0.048 & Boxy-peanut bulge and nuclear disc & 24.8\\     

\hline                           
\end{tabular}
\tablefoot{The properties in this table are as follows: Galaxy ID, global star formation rate (SFR) \citep[WISE W4 band, ][]{2014ApJ...782...90C}, dust proxy (median \update{E(B–V)}) \citep[\ngist,][]{2025A&A...700A.237F}, previously classified kinematic structure \citep{2025A&A...700A.237F}, and assumed distance from Earth \citep[CF4, Table 5, column 2,][]{2023ApJ...944...94T}.}
\end{table*}
 
\subsection{Data reduction and analysis}
Data reduction was performed using the \textsc{python} package \textsc{pymusepipe}\footnote{https://github.com/emsellem/pymusepipe} \citep{2022A&A...659A.191E}, and is described in \cite{2025A&A...700A.237F}. \textsc{pymusepipe} was used to create mosaicked data cubes from MUSE science exposures. \textsc{pymusepipe} was built around the MUSE Data Reduction Pipeline \citep{2020A&A...641A..28W} and behaves as a data organiser and wrapper for the ESO Recipe Execution Tool \citep[\textsc{esorex,}][]{esorex2015}.

To extract stellar kinematics from our data, we applied the \ngist\footnote{https://github.com/geckos-survey/ngist} pipeline \citep{2025A&A...700A.237F} on the fully reduced and mosaicked datacubes. \ngist\ is an upgraded version of the \gist\ pipeline \citep{2019A&A...628A.117B}, which is a wrapper for existing spectral fitting routines for the analysis of IFS galaxy data. \ngist\ is a publicly available, modular, documented\footnote{https://geckos-survey.github.io/gist-documentation/} code applicable to any galaxy IFS data.

In this study, we utilise \ngist\ version 7.2.1 to generate 2D maps of 2-moment stellar kinematics, applying Voronoi binning to achieve a signal-to-noise ratio (S/N) of 100 per pixel (equivalent to 80 \AA$^{-1}$, calculated over the wavelength range of 4800–7000 \AA). This wavelength range was chosen to minimise the inclusion of skylines in the S/N estimate. We chose a binning of S/N = 100 as we are focused on the central regions of our galaxies, where S/N is very high and spaxels typically exceed this requirement regardless. The kinematics were derived from a wavelength range of 4800–8700 \AA. The Jeans equations, as implemented in JAM, are a function of the true moments $V$ and $\sigma$. As a result, we approximate these true moments by enforcing a Gaussian LOSVD (fixing all high-order Gauss-Hermite coefficients to zero). 

We utilised the penalized Pixel Fitting (pPXF) routine, as described by \cite{2004PASP..116..138C} and \cite{2017MNRAS.466..798C} in the \ngist\ stellar kinematics (\texttt{KIN}) module, in combination with the X-shooter stellar template library \citep{2022A&A...660A..34V}. We chose to use the X-shooter stellar library for its broad wavelength coverage, high spectral resolution near the CaT, and agreement with prior studies suggesting that stellar spectra are generally more reliable than SSPs for determining stellar kinematics \citep[e.g.][]{2017ApJ...835..104V,2019AJ....158..160B}. Following previous IFS works including SAURON \citep{2004MNRAS.352..721E}, ATLAS$^{\text{3D}}$ \citep{2011MNRAS.413..813C}, SAMI \citep{2017ApJ...835..104V}, MaNGA \citep{2019AJ....158..160B,2019AJ....158..231W} and PHANGS \citep{2022A&A...659A.191E}, a Legendre polynomial is applied to better match the data to the spectral templates. Motivated by the analysis of \cite{2017ApJ...835..104V} and its application to the GECKOS dataset in \cite{2025A&A...700A.237F}, we chose a 23$^{\text{rd}}$-order additive Legendre polynomial. A first-order multiplicative polynomial is also fitted to correct minor continuum variations caused by imperfect sky subtraction and dust attenuation. Initial velocity guesses are sourced from the NASA Extragalactic Database, with a starting stellar velocity dispersion guess of 100 km s$^{-1}$. In this work, we used the nGIST output mean line-of-sight stellar velocity ($V$) and stellar velocity dispersion ($\sigma$) maps, along with the corresponding Voronoi bin positions. From the \ngist\ stellar populations and star formation histories module (\texttt{SFH}), we used the stellar dust absorption \update{E(B–V)} maps for our dust masking, calculated assuming a \cite{2000ApJ...533..682C} extinction curve. Additionally, we note that \ppxf has been shown to underestimate formal uncertainties \citep[e.g.][]{2019A&A...631A.130B,2025A&A...697A..94G}, with \cite{2019A&A...631A.130B} demonstrating that the true uncertainty in stellar velocity dispersion is $20\%$ higher than the value returned by \ppxf. Resultingly, we increase our $\sigma$ uncertainty by $20\%$, and conservatively increase our $V$ uncertainty by $20\%$ as well. As a final step, we calculated the median velocity from the $V$ map within a circular aperture of diameter 1 kpc, centred on the galaxy, and subtracted this systemic velocity from the velocity map.

\subsection{Surface brightness modelling}
Dynamical modelling with the JAM formalism requires a model for the gravitational potential (which can also include e.g., a dark matter halo), and the tracer stellar population. We thus require both a) imaging of our galaxy sample that accurately represents the stellar distribution (i.e. unimpacted by dust), and b) a method for modelling this stellar distribution which is efficient and allows for computationally efficient evaluation of the Jeans equations.
\subsubsection{Spitzer 3.6$\mu$m imaging}
\label{sec:imaging}
As our sample contains highly inclined galaxies, the impact of dust absorption along the dust lane/photometric major axis is strong and requires care with the selection of our photometry. Any imaging in a wavelength band similar to the kinematic maps wavelength range (e.g. SDSS-r band) will be strongly affected by dust. For this reason, we chose to use 3.6$\mu$m mid-IR imaging from the Spitzer Space Telescope \citep{2004ApJS..154...10F}. Cutouts of each galaxy in our sample were downloaded from the NASA IPAC Infrared Science Archive, and are shown in Figure \ref{fig:sample_imaging}.
\subsubsection{Multi-Gaussian expansion profiles}
\label{sec:mge}
The majority of previous work with JAM has utilised the MGE parametrisation of \cite{1992A&A...253..366M,1994A&A...285..723E}, which can accurately reproduce the surface brightness of real galaxies and has an efficient and widely-used routine \citep{2002MNRAS.333..400C}. This approach assumes the 2D projected luminosity, $I$, of a galaxy on the sky can be represented by a sum of N Gaussians. Each Gaussian has total luminosity $L_k$, an observed axial ratio $q'_k$ and a dispersion $\sigma_k$ along the major axis. This model may be then convolved with a second sum of Gaussians representing the PSF, which allows the model to stay within the MGE formalism.

\begin{figure}
    \centering
    \includegraphics[width=\linewidth]{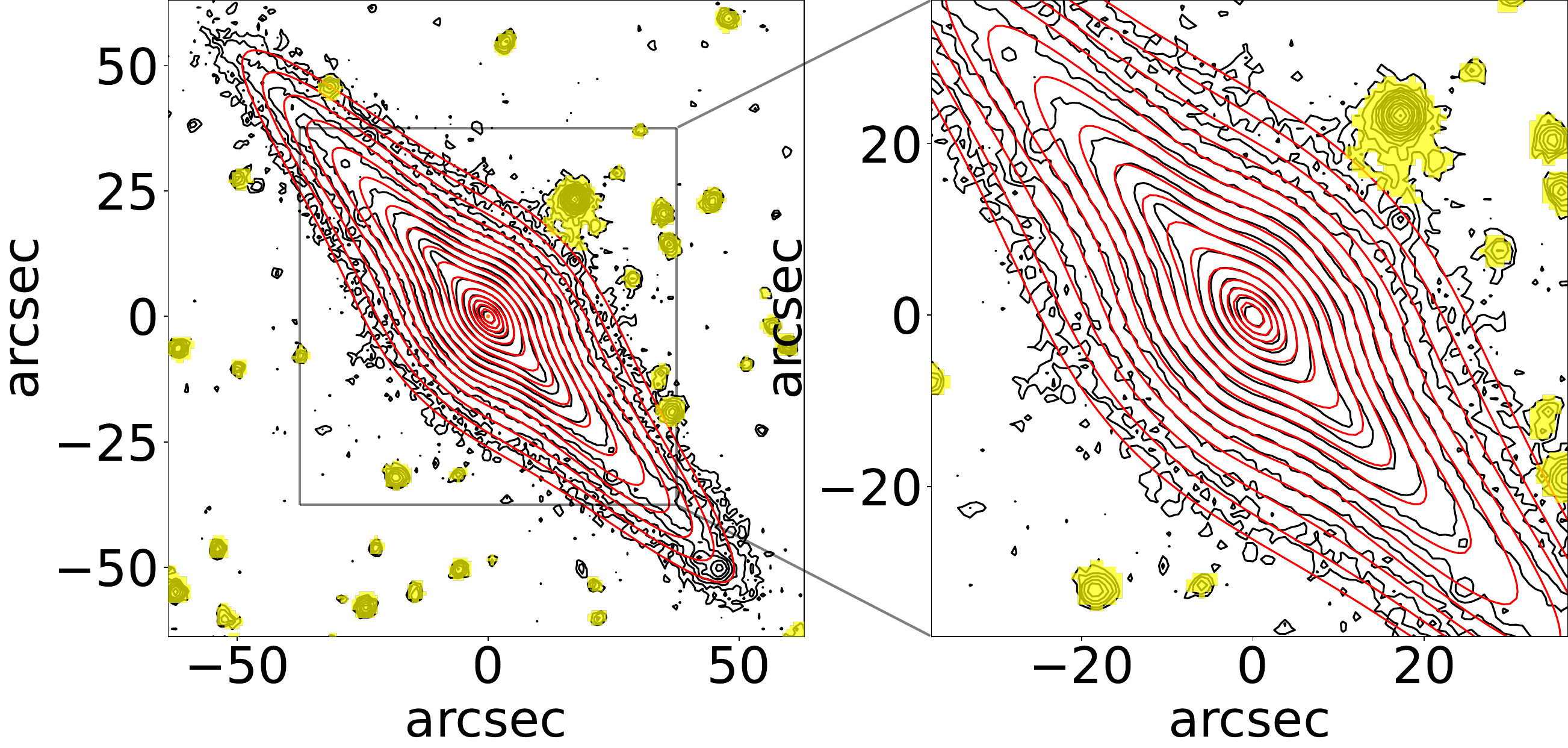}
    \caption{MGE fit to the galaxy IC\,1711. Spitzer 3.6$\mu$m imaging contours are in black, the MGE model contours are in red, and masked regions (other sources) are in yellow. The contours are spaced by 0.5 mag arcsec$^{-2}$. The right side panel is a zoomed-in version of the left side panel, with the scale shown by a box around the central region in the left side panel.}
    \label{fig:IC1711_mge}
\end{figure}

We create MGE models of the Spitzer 3.6$\mu$m imaging for each galaxy in our sample. We use a Gaussian with a FWHM of 1.66" \citep[the average PSF for IRAC 3.6$\mu$m imaging;][]{2017ApJS..228....5K} for our PSF. We fix the position angle of all Gaussians to be the same, as JAM cannot account for non-axisymmetric structure.
We show an example of an MGE fit to IC\,1711 in Figure \ref{fig:IC1711_mge}. Similar fits were created for all galaxies in the sample. These resulting Spitzer MGEs were used as the model for the tracer stellar population, and its contribution to the total gravitational potential.

\section{Methods}
\label{sec:methods}
We use the solutions to the Jeans equations \citep{1922MNRAS..82..122J} in cylindrical coordinates, using the JAM code of \cite{2008MNRAS.390...71C}, for a more detailed explanation of the formalism, see \cite{2008MNRAS.390...71C}.

In this work, we have assumed that the velocity ellipsoid is aligned with cylindrical co-ordinates, which has been shown to be accurate near the equatorial plane and along the minor axis of fast rotators \citep{2007MNRAS.379..418C}. We use our MGE model as the luminous density for JAM, and then multiply by a global and constant mass-to-light ratio and add a spherical NFW \citep{1997ApJ...490..493N} dark matter profile to obtain our gravitational potential. The NFW profile is defined by only one parameter, the dark matter fraction within $1R_e$. The NFW slope is set to $\gamma=-1$ as in \cite{1997ApJ...490..493N}, and the halo break radius to $r=20$~kpc. We do this as the halo break radius is not well constrained by typical spectroscopic data-sets, which are heavily biased towards the central baryon-dominated \update{regions (of MW-mass galaxies)}, and setting $r=20$~kpc provides equivalently good models as letting it vary \citep{2018MNRAS.476.4543B}. \update{Given that our GECKOS MUSE data does not extend beyond $r=20$~kpc, the precise choice of break radius should not significantly affect dynamical parameters, nor the velocity residuals at even smaller radii.}

We construct JAM models for the seven galaxies in our sample. We use the \textsc{emcee} package \citep{2013PASP..125..306F} in \textsc{python} to fit the free parameters of the model for each galaxy, where the likelihood function is simply the JAM procedure of \cite{2008MNRAS.390...71C} and its returned global $V_{\text{rms}}$ $\chi^2_{\text{reduced}}$. We use a uniform prior across a reasonable estimation range for each free parameter. These free parameters are stellar mass-to-light ratio, dark matter fraction, inclination, and $\beta_j$ the stellar orbital anisotropy of the individual Gaussians which compose the stellar MGE. JAM then outputs a $V_{\text{rms}}$ map, where $V_{\text{rms}}^2 = V^2+\sigma^2$, $V$ is the mean stellar velocity, and $\sigma$ is the stellar velocity dispersion. For each model, we apply a dust mask to the observed kinematic maps based on the E(B–V) stellar dust absorption maps, masking any Voronoi bin with E(B–V) above a given threshold. Our initial models adopt a cut-off value of E(B–V) = 0.7, but we also use several thresholds down to \update{E(B–V)} = 0.2 to test the robustness of our results against the impact of dust. Figure \ref{fig:masking} illustrates the difference between our highest and lowest masking thresholds, with \update{E(B–V)} > 0.7 masked in the left column, and \update{E(B–V)} > 0.2 masked in the right column. Finally, the first velocity moment, $V$, can be found by defining a $\kappa$ parameter for our luminous Gaussians. The $\kappa$ parameter defines the amount of rotation in $V_{\text{rms}}$. \update{Derived dynamical parameters should not be affected by our choice of $\kappa$, as they are only dependent on the $V_{\text{rms}}^2$ map. In principle, allowing $\kappa$ to vary across the luminous Gaussians could improve the recovery of the $V$ map, as different components such as thin discs, thick discs, and nuclear discs have different rotational support. However, as our goal is simplicity in the model, we adopt a single constant $\kappa$, scaled such that the model velocity field has the same projected angular momentum as our observed galaxy.}

\section{Results}
\label{sec:results}
\subsection{JAM models}

\begin{figure*}
    \centering
    \includegraphics[width=\linewidth]{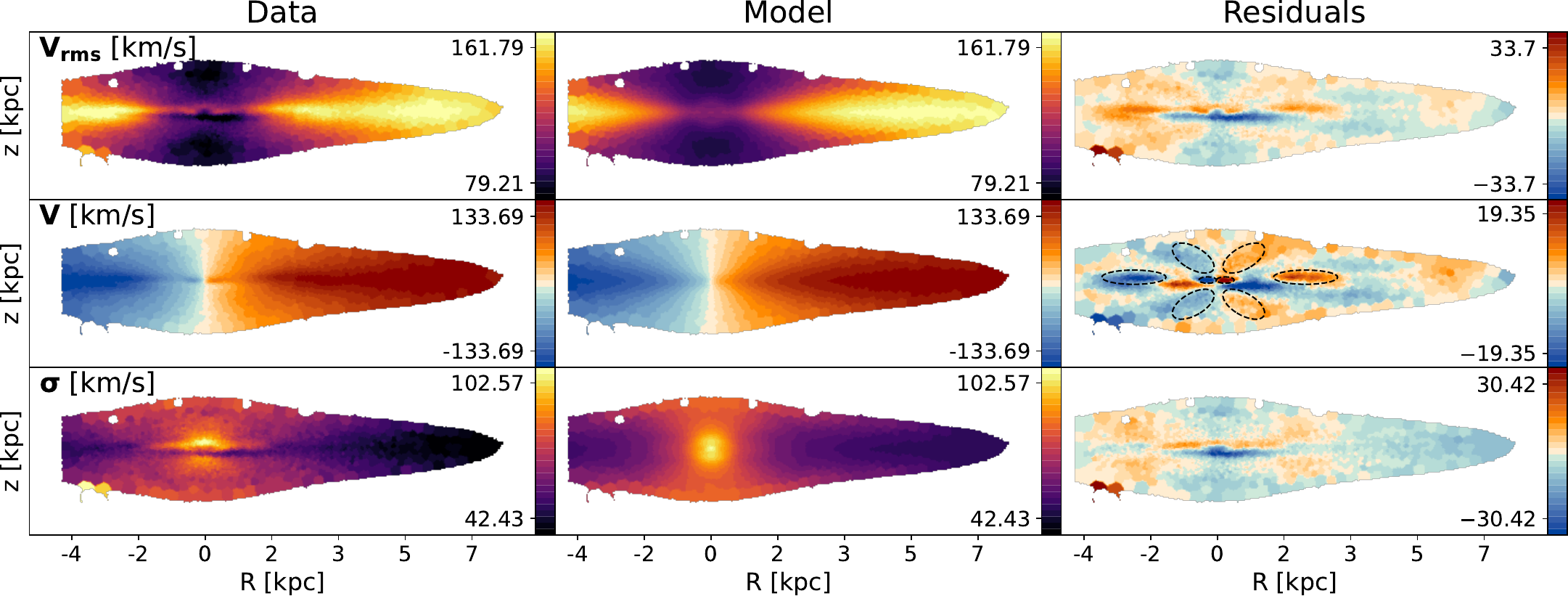}
    \caption{JAM model for galaxy NGC\,3957. The upper row shows the $V_{\text{rms}}$ map, the central row shows the $V$ map, and the lower row shows the $\sigma$ map. The left column shows \ngist\ mean light-weighted velocity $V$ and $\sigma$ binned to S/N=100, as well as the derived quantity $V_{\text{rms}}=\sqrt{V^2+\sigma^2}$. The central column shows the dynamical model, and the right column shows the residuals (data-model). The model is fit to the $V_{\text{rms}}$ map, and a $\kappa$ value is fit to find the $V$ map from $V_{\text{rms}}$. Additionally, we circle structure in the $V$ residual map that we believe corresponds to kinematic components that JAM has failed to successfully model.}
    \label{fig:NGC3957}
\end{figure*}

\begin{figure*}
    \centering
    \includegraphics[width=\linewidth]{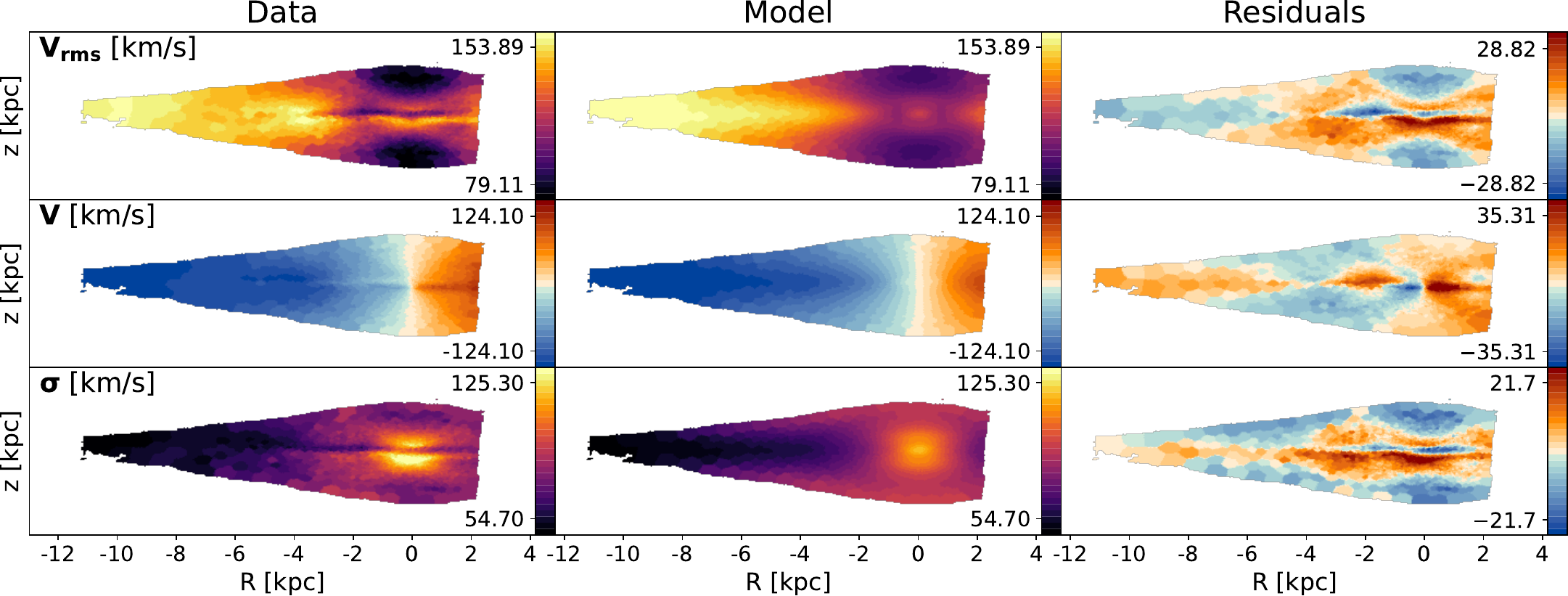}
    \centering
    \caption{As for Figure \ref{fig:NGC3957}, but for IC\,1711, and without circling structure.}
    \label{fig:IC1711}
\end{figure*}

\begin{figure*}
    \centering
    \includegraphics[width=\linewidth]{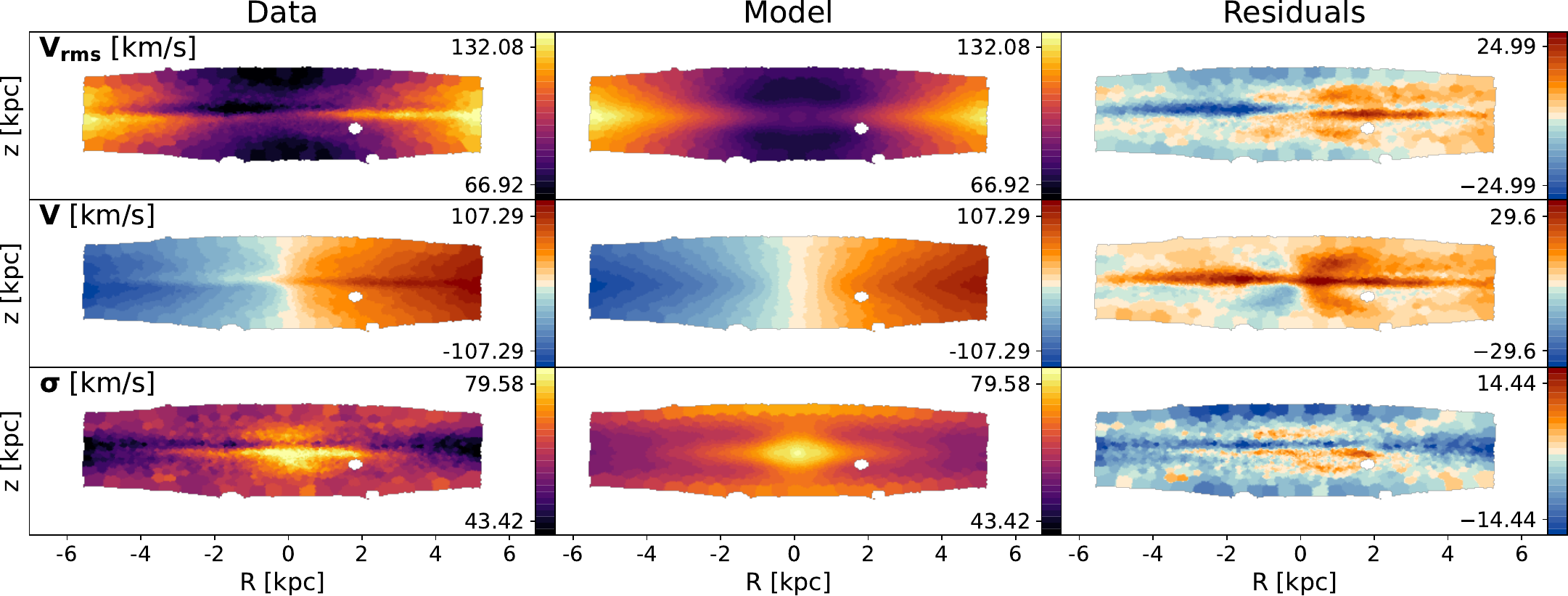}
    \caption{As for Figure \ref{fig:NGC3957}, but for NGC\,0522, and without circling structure.}
    \label{fig:NGC0522}
\end{figure*}

We fit JAM models to each galaxy in our sample, using our \update{E(B–V)} < 0.7 requirement for each Voronoi bin in this first set of models. Maps of $V_{\text{rms}}$, and the derived quantities $V$ and $\sigma$, are shown in Figures \ref{fig:NGC3957}-\ref{fig:NGC0522} for NGC\,3957, IC\,1711 and NGC\,0522. The remaining galaxies do not exhibit clear non-axisymmetric structure, and their maps are shown in the appendix, in Figures \ref{fig:NGC3279}-\ref{fig:UGC00903}. The left column shows $V_{\text{rms}}$, $V$ and $\sigma$ derived from \ngist\ output for the seven GECKOS galaxies, the central column shows the same but from the JAM model, and the right column shows the residuals, i.e. data minus model. 

Measuring the goodness of fit of our models is not straightforward, particularly given that non-axisymmetric structures are expected to induce systematic discrepancies between the observed and modelled kinematics, particularly in the bulge-dominated regions of our galaxies, and one of our aims is to use these discrepancies where possible to diagnose these structures. As a result, instead of evaluating $\chi^2_{\text{reduced}}$ as a global statistic, we evaluate it as a function of radius. We compute $\chi^2_{\text{reduced}}$ as a moving average, within radial bins of $0.5\times R_d$, where $R_d$ is the disc scale radius \citep{2015ApJS..219....4S}\footnote{Disc scale radii for Spitzer 3.6~$\mu$m imaging were derived for multi-component S\'ersic fits. We estimated $R_d$ for NGC\,3957 from a 2-component S\'ersic fit to its photometry, as this was not calculated by \cite{2015ApJS..219....4S}.}. This radial smoothing allows us to identify how well the different regions of our galaxy (e.g. bulge dominated, disc dominated) are fit.

\begin{figure*}
    \centering
    \includegraphics[width=\linewidth]{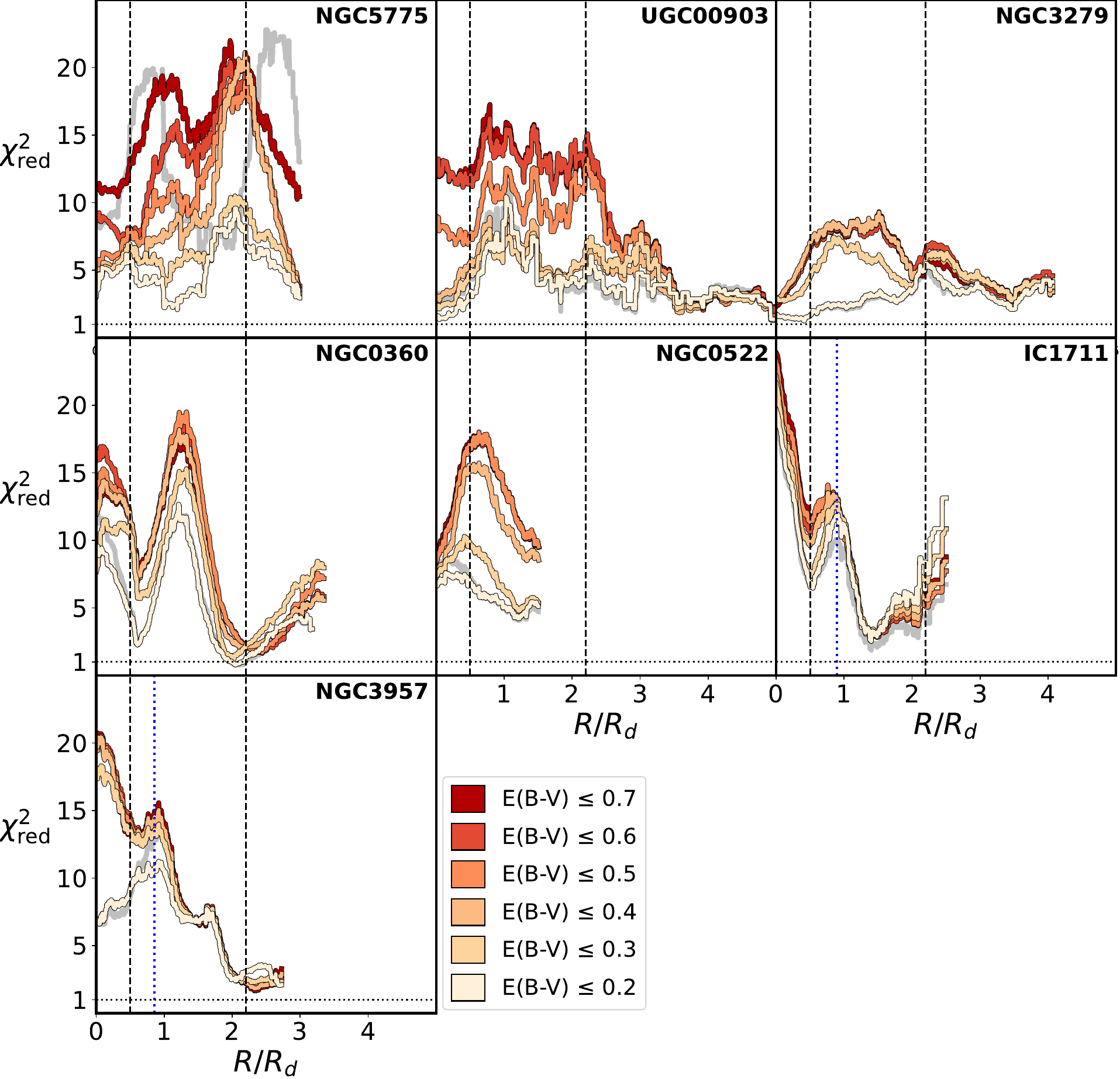}
    \caption{The moving average $\chi^2_{\text{reduced}}$ value for each galaxy in our sample, plotted as a function of $R/R_d$, where $R_d$ is the disc scale length. For each galaxy, and each dust masking cut-off, we measure $\chi^2_{\text{reduced}}$ in a $0.5\times R_d$ bin, spaced in radius along the major axis. $\chi^2_{\text{reduced}}=1$ is shown as a dotted line, and $R_d=0.5$ (bulge dominated region, \citealp[e.g.][]{2010ApJ...716..942F}) and $R_d=2.2$ (disc dominated region, \citealp[e.g.][]{1970ApJ...160..811F,1996MNRAS.281...27P}) as vertical dashed lines. We also plot in grey the $\chi^2_{\text{reduced}}$ measure for our \update{E(B–V)}~$\leq$~0.7 model, but applied to \update{E(B–V)}~$\leq$~0.2 bins. Blue dotted lines are plotted at 3.5~kpc for IC\,1711 and 2.5~kpc for NGC\,3957, the radius where their photometric shoulders end (see Section \ref{sec:NGC3957_analysis}). We note that NGC\,5775 and NGC\,0360 are not perfectly edge-on, and show clear spiral arms in their photometry, which explains their strong peaks in $\chi^2_{\text{reduced}}$. A general trend of decreasing $\chi^2_{\text{reduced}}$ with mask level is seen, as well as discs being fit better than bulges.}
    \label{fig:rd_measure}
\end{figure*}

Figure \ref{fig:rd_measure} illustrates the goodness of fit for each galaxy in our sample, as a function of radius and dust masking level. In each panel, we show the moving average of $\chi^2_{\text{reduced}}$, for each mask threshold in \update{E(B–V)}. The $x$-axis is shown as a function of disc scale radius, and the $y$-axis has the same range for each galaxy, to aid in comparison. We show $\chi^2_{\text{reduced}}=1$ as a dotted line, and $R_d=0.5$ (bulge dominated region \citep[e.g.][]{2010ApJ...716..942F}) and $R_d=2.2$ (disc dominated region \citep[e.g.][]{1970ApJ...160..811F,1996MNRAS.281...27P}) as vertical dashed lines. Finally, we plot the $\chi^2_{\text{reduced}}$ measure for our \update{E(B–V)}~$\leq$~0.7 model, but only applied to \update{E(B–V)}~$\leq$~0.2 bins in grey. We firstly find that in the disc dominated region ($0.5<R/R_d<2.2$), most galaxies approach $\chi^2_{\text{reduced}}\lessapprox5$, indicating a reasonable fit for the disc component of the galaxy, in comparison to the bulge. In our most star forming galaxies (NGC\,5775, UGC\,00903, NGC\,3279) there is a strong improvement with more aggressive dust masking, with this being less apparent but still true in the bulge dominated region ($R/R_d<0.5$) for NGC\,0360 and NGC\,0522. However, IC\,1711 and NGC\,3957 show little improvement with stronger masking. Interestingly, these galaxies have previously been identified as hosting non-axisymmetric structure \citep[bars,][]{2025A&A...700A.237F}.This could hence be indicative of these non-improving galaxies having non-axisymmetric structure, but these galaxies also have the lowest SFR and therefore presumed dust content, and hence we may just be seeing that masking fewer bins gives a smaller improvement. The $\chi^2_{\text{reduced}}$ measure for our \update{E(B–V)}~$\leq$~0.7 model applied to \update{E(B–V)}~$\leq$~0.2 bins generally follows the \update{E(B–V)}~$\leq$~0.2 model for most galaxies, with the exception of NGC\,5775. This is because for most galaxies, there aren't enough dusty bins with 0.2~<~\update{E(B–V)}~\update{<~}0.7 to affect the fit significantly, and $\chi^2_{\text{reduced}}$ is driven almost entirely by only measuring dust-free bins. In the case of NGC\,5775, there are enough dusty bins to affect the \update{E(B–V)}~$\leq$~0.7 model. Overall, in the disc dominated region ($0.5<R/R_d<2.2$), we find that most galaxies approach $\chi^2_{\text{reduced}}\lessapprox5$, indicating a reasonable fit for the disc component of the galaxy, in comparison to the bulge.

\begin{figure}
    \centering
    \includegraphics[width=\linewidth]{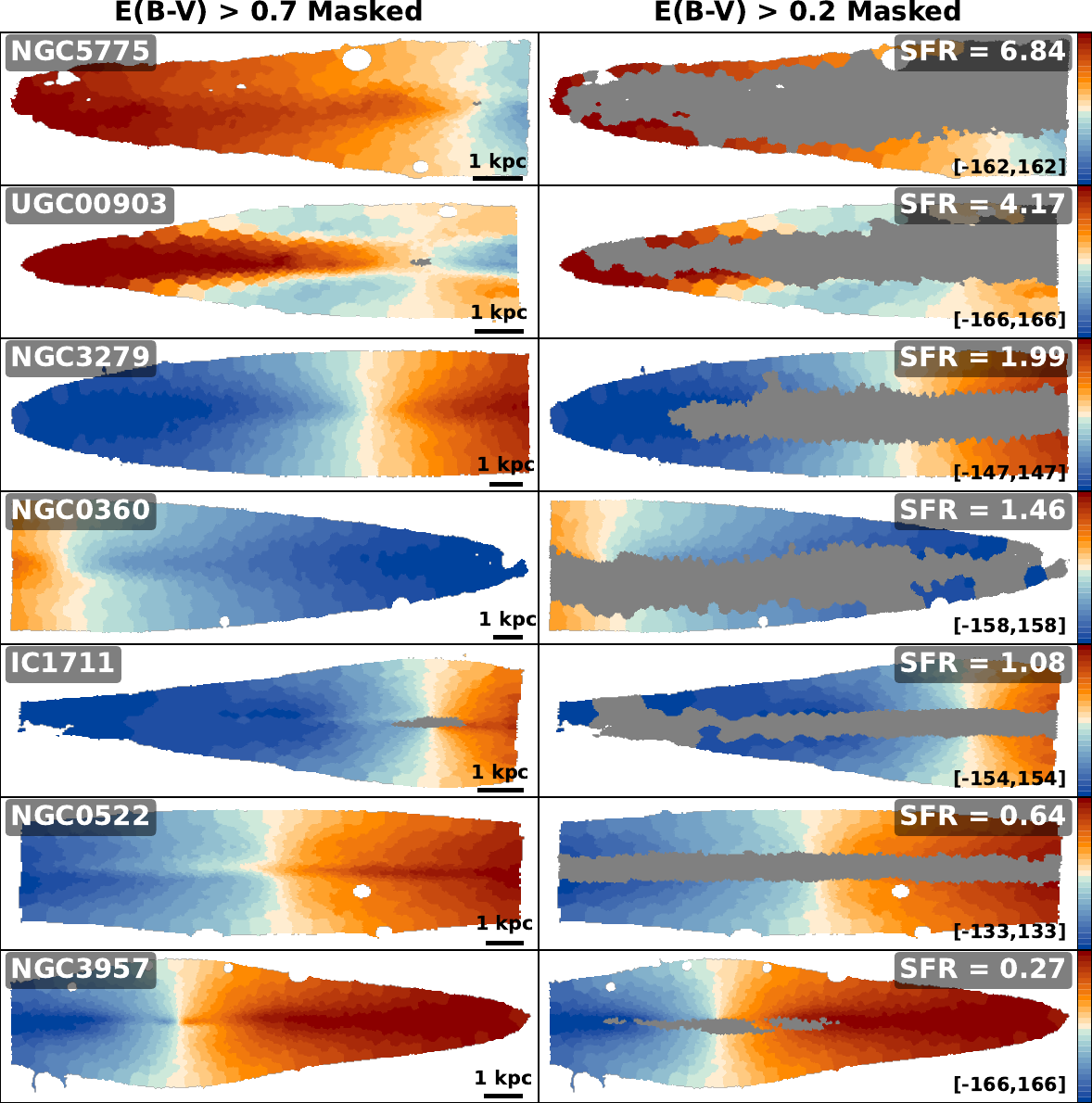}
    \caption{Illustration of masking for two E(B–V) thresholds across our sample, overlaid on the $V$ maps. Each row corresponds to a galaxy, ordered from top to bottom by decreasing star formation rate. The left column shows masks where Voronoi bins with \update{E(B–V)} > 0.7 are excluded (greyed out), while the right column shows masks for \update{E(B–V)} > 0.2. Galaxies are also labelled with their star formation rates in $\rm{M}_\odot~\text{yr}^{-1}$. We show the range in the velocity colourbar in the bottom right of the right-side panels. Aggressive masking removes most mid-plane bins in highly star-forming galaxies.}
    \label{fig:masking}
\end{figure}
 
\subsection{Kinematic residual map structure}
\label{sec:kinematic_structure}
\begin{figure*}
    \centering
    \includegraphics[width=\linewidth]{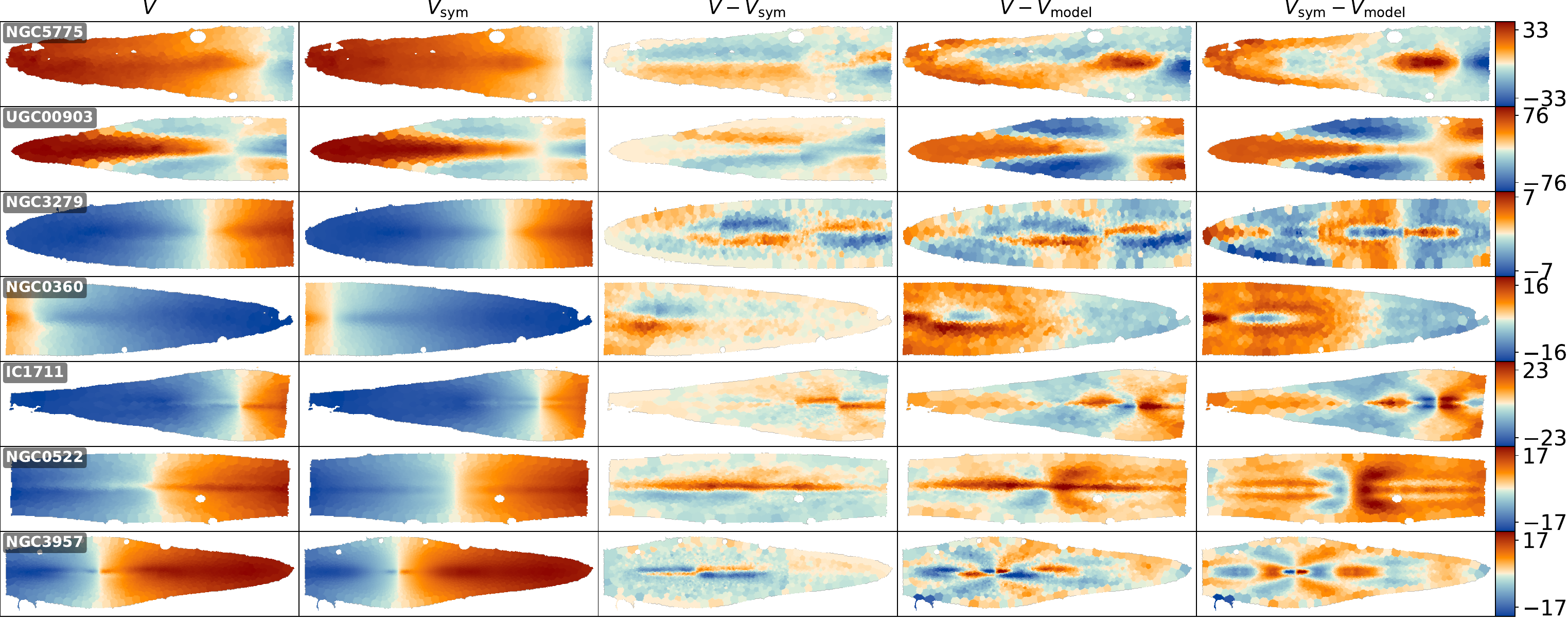}
    \caption{Velocity maps for all seven galaxies in our sample, demonstrating the effect of symmetrisation within JAM. The leftmost column shows the observed velocity ($V$) maps. The $V_{\text{sym}}$ column presents the same velocity maps after applying the symmetrisation procedure used by JAM. The $V - V_{\text{sym}}$ column shows the residuals between the observed and symmetrised velocities. The $V - V_{\text{model}}$ column shows the residuals between the observed velocity and the best-fitting JAM model (with a mask of \update{E(B–V)}<0.7). Finally, The $V_{\text{sym}} - V_{\text{model}}$ column shows the residuals between the symmetrised velocity field and the JAM model. Our diagnosis of non-axisymmetric velocity structures will consider only features that are also clearly visible in the $V_{\text{sym}} - V_{\text{model}}$ residuals, as these are unaffected by the symmetrisation process.}
    \label{fig:symmetrised_velocities}
\end{figure*}

We now examine the spatial distribution of kinematic residuals. These residual maps contain valuable information about localised deviations from axisymmetry, but we must be careful to take into consideration the impact of dust and masking. We note here that JAM fits to a \update{symmetrised} transformation of the input kinematics. Thus, some velocity residual structures are introduced purely by this symmetrisation process and should not be misinterpreted as non-axisymmetric velocity components. To illustrate this, in Figure \ref{fig:symmetrised_velocities} we show the $V$ maps for our sample, as well as the symmetrised $V_{\text{sym}}$ maps. We also compare the residuals $V-V_{\text{sym}}$, $V-V_{\text{model}}$ and $V_{\text{sym}}-V_{\text{model}}$. The $V_{\text{sym}}-V_{\text{model}}$ residual should be free of any structures due to the symmetrisation, and thus we will only classify structure that is clearly visible in this and the $V-V_{\text{model}}$ residual.

We begin our analysis with NGC\,3957, which has the lowest SFR, lowest median \update{E(B–V)}, and the best fit model with only small changes when we change our masking threshold. We thus expect that the JAM models of NGC\,3957 should have the smallest impact from dust obscuration, with any deviations from a good model fit due to non-axisymmetric structure, rather than dust. However, we take care to only consider structures in the kinematic residuals that are not impacted by dust (i.e. visible even with \update{E(B–V)}~<~0.2 masking applied, see Figure \ref{fig:masking}). Finally, we have the kinematic structure classification from \cite{2025A&A...700A.237F}, which suggests that NGC\,3957 has a BP bulge and nuclear disc. In Figure \ref{fig:NGC3957}, we highlight coherent structures in the $V$ residual maps of NGC\,3957, by circling these structures in dashed ellipses.

A compact, symmetric (around $R=0$) feature of higher residual velocity than its surroundings is apparent in the $V$ residual map ($R\lessapprox1$ kpc), suggestive of a nuclear disc. This residual feature peaks at approximately $~0.25$kpc along the major axis, in agreement with the size of the nuclear disc in NGC\,3957 found by \cite{2025A&A...700A.237F} (0.28kpc). Although nuclear discs are to a good approximation axisymmetric (although some host nuclear bars) and could, in principle, be modelled with JAM, accurate recovery would require finely tuned low-dispersion, highly flattened Gaussians in the MGE. We acknowledge here that there is a larger structure below the mid-plane with inverted signs at a similar scale, which we have not circled. However, it can be seen from Figure \ref{fig:symmetrised_velocities} that this is an artefact of the velocity symmetrisation process of JAM, so we do not consider it correlated with any physical structure.

At intermediate radii ($\sim2.5~$kpc), we identify and circle prominent residuals in the $V$ residual map along the major axis that stand out in magnitude and structure, exhibiting opposite signs across the disc and exceeding the surrounding residual levels. These structures suggest an excess of mean velocity, a few kpc along the major axis of the disc, on both sides of the galaxy centre. This is consistent with the kinematic signatures expected from a bar, where non-circular motions generate excess line-of-sight velocities near the bar ends \citep[e.g.][]{1992MNRAS.259..345A,1993RPPh...56..173S,2017A&A...606A..47F,2024ApJ...976..220K}.

Finally, the $V$ residual maps reveal an X-shaped structure. Such features are known indicators of BP bulges when observed photometrically via unsharp masking \citep[e.g.,][]{2006MNRAS.370..753B,2025A&A...700A.237F}, tracing stars on vertically resonant orbits. The detection of an X-shaped residual in the stellar kinematic $V$ map supports the interpretation of a BP bulge, as velocity moments can trace this structure off the plane of the disc \citep{2015MNRAS.450.2514I,2017A&A...606A..47F}, due to the bifurcated $x_1$ orbits that populate BP bulges \citep[e.g.][]{2002MNRAS.333..847S,2006MNRAS.366.1121P}.

We use NGC\,3957 as an illustrative example, given its clear kinematic signatures and known structures. The remaining galaxies with notable residuals are IC\,1711 and NGC\,0522, and are discussed in Section \ref{sec:extending_lessons}. The remaining galaxies in our sample (NGC\,0360, NGC\,3279, NGC\,5775, UGC\,00903) do not show residuals consistent with known structures, and hence we do not discuss their residuals in detail.

\section{Discussion}
\label{sec:discussion}
\subsection{The impact of dust on JAM models of edge-on disc galaxies}

Most previous applications of JAM modelling have been applied to passive, dust-free, and unbarred galaxies, where the model assumption of axisymmetry is likely to hold. Early work with JAM \citep[e.g.][]{1992A&A...253..366M,1994A&A...285..723E,1994A&A...285..739E,1999MNRAS.303..495E,2008MNRAS.390...71C,2009ApJ...704L..34C,2013MNRAS.432.1709C} was primarily performed on early-type galaxies from the SAURON and ATLAS$^{\text{3D}}$ surveys. These studies demonstrated that JAM models could robustly recover M/L, velocity anisotropy and dark matter fraction for elliptical and lenticular galaxies. There has been work done on barred systems \citep[e.g.][]{2012MNRAS.424.1495L,2016MNRAS.455.3680L}, but these have mostly been simulation works.

In contrast, dusty, star-forming disc galaxies pose significant challenges to dynamical modelling. These systems are affected by prominent dust lanes that distort both the observed light distribution and the weighting of the stellar kinematics. As a result, they violate JAM assumptions and are generally under-represented in the JAM literature. While some work has been carried out on JAM's reliability in more complex systems \citep[e.g.][]{2012MNRAS.424.1495L,2016MNRAS.455.3680L}, observational applications to highly spatially resolved, dusty, star-forming discs remain rare. Some work has been done on samples that contain spiral galaxies \citep[e.g.][]{2015MNRAS.451.2723S,2023MNRAS.522.6326Z}, but a strong discussion of the impact of dust was not included. This study therefore, represents one of the first applications of JAM to such galaxies with an intention to study the effects of dust. By focusing on a small but representative sample of edge-on discs with a range of dust distributions and star-formation rates, we explore whether JAM can still provide reliable dynamical constraints under these more difficult conditions.

To test whether reliable dynamical parameters could be extracted from partially masked data, we compared derived enclosed masses at 2.5kpc, 4.0kpc, and 10.0kpc, as well as derived inclination, across a range of dust masks from \update{E(B–V)} > 0.2 to \update{E(B–V)} > 1.0. The enclosed mass values are a function of the derived M/L and dark matter fraction, as the MGE light model for each galaxy remains consistent. All galaxies showed stable values within 10\%, indicating that global dynamical properties remain robust even when excluding the kinematics of the dust-dominated central regions. Interestingly, UGC\,00903 has both a comparable $V_{\text{rms}}$ fit and reliably determined dynamical parameters relative to the rest of the sample, despite hosting the most unusual kinematic structure, with a clearly visible counter-rotating thick disc (see Figure \ref{fig:UGC00903}). This is because dynamical parameters are recovered from just the $V_{\text{rms}}$ field, which is independent of how the kinematics separate into ordered and random motions, has no dependence on the sign of $V$, and makes no assumptions on the anisotropy. Indeed, JAM has been used previously to find the mass distribution of galaxies with counter-rotating discs \citep{2017MNRAS.464.4789M}.

Our results are consistent with previous simulation-based studies demonstrating that global M/L ratios in edge-on discs can be reliably recovered with JAM modelling, even in the presence of non-axisymmetric structures \citep{2012MNRAS.424.1495L}, and that the enclosed mass can be recovered to within 10\% when the true mass distribution is known \citep{2016MNRAS.455.3680L}. We extend these findings by showing that, although the true enclosed mass is unknown for our galaxies, the recovered mass remains consistent within 10\%, and is not biased across a wide range of dust masking thresholds.

\subsection{Connecting residuals to physical structure}
\subsubsection{NGC\,3957 as a benchmark system}
\label{sec:NGC3957_analysis}
NGC\,3957 provides a particularly clean case for interpreting residuals, due to its combination of low dust content, low star formation rate, and clear photometric features suggestive of non-axisymmetry. Most importantly, we are more confident that these features are due to non-axisymmetries, as the dust content is low (Figure \ref{fig:masking}), there is little change in our model when we apply more aggressive masking (Figure \ref{fig:rd_measure}), and the features are not apparent in the symmetrised velocity residuals (Figure \ref{fig:symmetrised_velocities}). As our best-fitting system, it provides an ideal benchmark for interpreting residuals in terms of physical substructures.

\begin{figure}
    \centering
    \includegraphics[width=\linewidth]{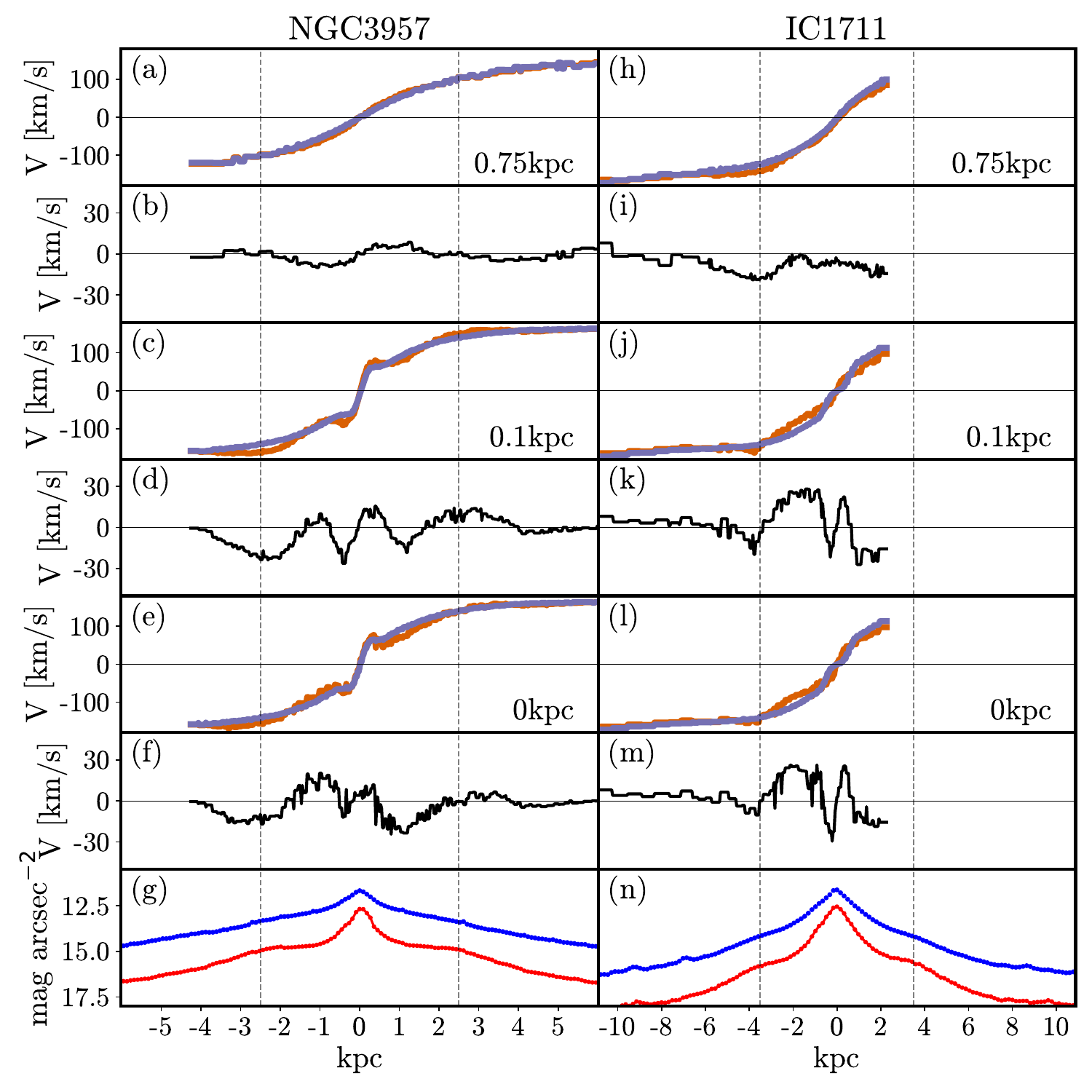}
    \caption{Stellar velocities and surface brightness profiles for NGC\,3957 and IC\,1711. For NGC\,3957, panels (a), (c) and (e) show velocities extracted from horizontal slits parallel to the major axis, with the data in orange, and the model in purple. Panel (a) shows a slit offset from the major axis by 0.75~kpc, panel (c) by 0.1~kpc, and panel (e) by 0~kpc. Panels (b), (d) and (f) show the residual velocity from the slits in the panels above, (a), (c) and (\update{e}) respectively. Panel (g) shows the surface brightness profiles from Spitzer IRAC 3.6~$\mu$m imaging. The total flux summed along the minor axis for each point along the major axis is shown in blue, and just the flux along the major axis is shown in red. There are dashed black vertical lines at $\pm2.5$~kpc, indicating where the photometric shoulders in the major axis profile end. For IC\,1711, we have the same, but panels (h), (j), and (l) show the velocities; (i), (k) and (m) show the residuals, and panel (n) shows the surface brightness profiles. There are dashed black vertical lines shown at 3.5~kpc, where the photometric shoulders end.}
    \label{fig:NGC3957_barlength}
\end{figure}

Photometric evidence supports the presence of a bar in NGC\,3957. As discussed by \cite{2005ApJ...626..159B}, Freeman Type II surface brightness profiles \citep{1970ApJ...160..811F} are suggestive evidence of bar structure \citep[e.g.][]{2003ApJ...583L..75G,2008AJ....135...20E,2016MNRAS.462.3430K}, due to resonances and instabilities in the bar leading to a redistribution of disc material. A Freeman Type II profile occurs when the major-axis surface brightness exhibits a local depression beyond the central component, followed by a flat or slightly rising plateau (depending on the bar orientation) before transitioning into an outer exponential decline. Of the seven GECKOS galaxies in our sample, only NGC\,3957 and IC\,1711 clearly display this shoulder in their major-axis 3.6$\mu$m surface brightness profiles, which we show in Figure \ref{fig:NGC3957_barlength}. We also note that NGC\,0522 shows possible kinematic evidence for a bar (see Section \ref{sec:extending_lessons}) and its surface brightness profile shows noticeable flattening, but less clear shoulders than NGC\,3957 and IC\,1711. We caution that this signature is only visible when the bar is not oriented end-on, as projection effects can obscure the shoulder \citep[e.g.][]{2000A&A...362..435L,2005MNRAS.358.1477A}.

In Section \ref{sec:kinematic_structure}, we highlighted the different structures visible in the $V$ map residuals for NGC\,3957. Here, we attempt to correlate these velocity residuals with photometric evidence for a bar. If a bar is present, we expect non-axisymmetric stellar motions, particularly those from elongated $x_1$ orbits in the disc plane, to manifest as velocity excesses along the disc major axis. For this, we extract velocities from the data, model, and residual maps along three slices of 1.5" thick: one along the mid-plane (0 kpc offset), one just above it (0.1 kpc), and one further off-plane  (0.75 kpc). All slices were taken on one side of the mid-plane, on the opposite side of the dust lane. This allows us to probe both in-plane structures, and off-plane structures away from the dust lane.

In Figure \ref{fig:NGC3957_barlength} we show the velocity profiles along these slits, for NGC\,3957 and IC\,1711, with the data represented by the orange line, the model by the purple line, and the residual by the black line. We also show the photometric shoulders from Spitzer IRAC 3.6~$\mu$m imaging. The vertical dashed lines are drawn to approximately where the photometric shoulders end. For NGC\,3957, this corresponds to a radius of $\sim 2.5$kpc, almost exactly where the peak is in the residual velocities, seen most strongly in the 0.1~kpc offset slit. 

We include IC\,1711 in the same figure for comparison. Although this galaxy has more dust, it also exhibits a Freeman Type II profile and strong velocity residuals. The photometric shoulders for IC\,1711 are at a radius of $\sim 3.5$kpc, corresponding almost exactly to where the excess velocities drop to zero after a peak, seen most strongly in the 0kpc and 0.1kpc offset slits. While this is different to NGC\,3957 where the shoulders spatially corresponded to a peak in velocity residuals, the resemblance to NGC\,3957 still lends further support to the interpretation that such kinematic structures can trace bar-like components, even in dustier systems. Further to this, we see strong peaks in the $V_{\text{rms}}$ $\chi^2_{\text{reduced}}$ at 2.5~kpc for NGC\,3957 and 3.5~kpc for IC\,1711 in Figure \ref{fig:rd_measure}. As modelled $V_{\text{rms}}$ requires \update{fewer} assumptions to derive than modelled $V$, this lends stronger weight to the correlation between these kinematic residuals and the presence of a bar. We emphasise here that we are simply showing that there is a spatial correlation between the kinematic residuals and photometric major-axis shoulders in both NGC\,3957 and IC\,1711. We do not make predictions on what orbital structures cause this, nor any implied bar size or position angle.

\subsubsection{Extending lessons from NGC\,3957 to our remaining sample}
\label{sec:extending_lessons}
Building on the lessons from NGC\,3957, we explore whether similar interpretations of residuals can be extended to dustier, more star-forming galaxies in our sample. We note here that NGC\,3957 has been identified as hosting a BP bulge in previous works \citep{1999AJ....118..126B,2000A&A...362..435L,2025A&A...700A.237F}, as well as IC\,1711 \citep{2015ApJS..217...32B,2025A&A...700A.237F} and NGC\,0522 \citep{2000A&A...362..435L,2015ApJS..217...32B,2025A&A...700A.237F}.
While increased dust complicates the analysis, we suggest several possible structures due to non-axisymmetric orbits in the residual maps of galaxies IC\,1711 and NGC\,0522.

IC\,1711 (Figure \ref{fig:IC1711}) has already been shown to share several key properties with NGC\,3957 in Section \ref{sec:NGC3957_analysis}: both exhibit clear Freeman Type II surface brightness profiles (Figure \ref{fig:NGC3957_barlength}) and coherent residual structures in their velocity maps. Here, we note that the major axis residuals in IC\,1711 are indeed still somewhat visible even with our most aggressive mask (E(B–V)~>~0.2). This supports the interpretation that, despite IC\,1711's higher dust content (median \update{E(B–V)} of 0.107, c.f. 0.048 for NGC\,3957), its kinematic residuals are still spatially correlated with photometric evidence for a bar, and could be tracing bar orbits.

NGC\,0522 presents a more ambiguous case. Unlike NGC\,3957 and IC\,1711, it does not exhibit a clear Freeman Type II profile, though it does show noticeable flattening in its major-axis surface brightness distribution. The velocity residuals in NGC\,0522 (Figure \ref{fig:NGC0522}) also lack the distinct major-axis excess seen in NGC\,3957 and IC\,1711. However, they do show a notable X-shaped pattern in the off-plane regions of the residual map, qualitatively similar to the pattern we previously noted for NGC\,3957 in Section \ref{sec:kinematic_structure}. Importantly for this analysis, this pattern is still visible regardless of our dust mask. This structure is morphologically similar to the residuals produced by BP bulges: vertically thickened inner bar structures that arise from dynamical instabilities \citep{1981A&A....96..164C, 2005ApJ...626..159B}. Indeed, \cite{2014MNRAS.444L..80L} and \cite{2025A&A...700A.237F} found that NGC\,0522 shows this structure in unsharp-masked 3.6~$\mu$m imaging. Buckled BP bulges have been shown to induce complex non-circular motions, especially in edge-on projections. \cite{2015MNRAS.450.2514I} found that the higher-order $h_3$ and $h_4$ maps trace BP off the plane of the disc, and \cite{2017A&A...606A..47F} found the same for mean velocities. While we cannot conclusively identify a bar in NGC\,0522, the presence of this X-shaped kinematic signature, as well as the flattened major-axis surface brightness distribution, provides evidence for a BP bulge. Given that previous work has established the presence of a BP bulge in imaging, the lack of strong kinematic evidence for a bar implies a side-on bar orientation, which is the same conclusion reached by \cite{2025A&A...700A.237F} for this galaxy.

\subsubsection{General trends across the sample and practical guidelines}

While we focus on detailed residual interpretation for three systems, Figure \ref{fig:rd_measure} shows that all galaxies in our sample reach acceptable JAM disc fits, with dust masking important for our most star-forming galaxies (NGC\,5775, UGC\,00903, NGC\,3279), consistent with the expected correlation between dust content and SFR in star-forming discs \citep[e.g.][]{2001PASP..113.1449C,2019ApJ...882...65M,2021A&A...655A.104P,2022MNRAS.513.2904T}. Despite this, derived enclosed mass remains consistent across mask levels, with variations below 10\%, in line with earlier JAM analyses \citep{2012MNRAS.424.1495L,2016MNRAS.455.3680L}.

Crucially, velocity residuals still reveal coherent kinematic signatures of non-axisymmetric structures in NGC\,3957, IC\,1711, and NGC\,0522 even when an aggressive \update{E(B–V)} > 0.2 mask is applied, and masked bins are not considered. These detections are supported, in the case of NGC\,3957 and IC\,1711, by photometric evidence for a bar. We suggest that residual-based identification of internal structure remains viable under substantial dust masking, provided the features persist.

We therefore acknowledge that applying a dust mask of at least \update{E(B–V)}~>~0.7 is sufficient to recover reliable dynamical parameters in edge-on systems. However, given that these derived dynamical parameters are consistent across dust maskings, and residual-based identification of non-axisymmetric structure is most reliable when a strong dust mask is applied, we recommend applying a dust mask of \update{E(B–V)}~>~0.2 for general application of JAM models to edge-on disc galaxies.

\section{Summary and conclusions}
In this paper, we investigated the limitations of axisymmetric dynamical modelling of edge-on disc galaxies, focusing on the combined effects of dust attenuation and non-axisymmetric structure on stellar kinematic residuals. We constructed Jeans Anisotropic MGE (JAM) models for seven galaxies in the GECKOS VLT/MUSE survey, using 3.6$\mu$m Spitzer IRAC photometry for the light and mass model, with an additional NFW dark matter profile included in the mass model.

One of the goals of this work was to assess what information about non-axisymmetric kinematic structures could be recovered from edge-on galaxies by subtracting a simple, axisymmetric dynamical model. By creating JAM models of a sample of GECKOS galaxies with varying dust content and structural complexity, we test whether coherent features in the velocity residuals could reveal underlying non-axisymmetric structure. The velocity residuals between data and model in NGC\,3957, our least dusty target, reveal coherent patterns aligned with expected bar orbits and photometry, suggesting a clear link between residual structure and underlying non-axisymmetric kinematics. We extend this analysis to dustier galaxies which shows similar velocity residuals to NGC\,3957. IC\,1711 shows similar coherent patterns aligned with photometry, despite the impact of dust. NGC\,0522 also has residuals strongly impacted by dust, but off-plane structures show promise of a diagnostic of vertical bar instabilities. Our results therefore suggest that residual maps from JAM can be a powerful diagnostic for barred structure, provided that regions with dust extinctions of \update{E(B–V)}~>~0.2\update{ are masked}.

We find that JAM fits discs well in all galaxies, and applying stricter E(B–V) masks (e.g. $>$ 0.2 or $>$ 0.4) resulted in $\chi^2_{\text{reduced}}\leq5$ in the disc region. Additionally, all galaxies showed consistent (within 10\%) values of enclosed mass and inclination across a range of dust masks. Notably, galaxies previously classified as containing central non-axisymmetric structure by \cite{2025A&A...700A.237F} showed the smallest changes with increased masking. This is likely due to a combination of these galaxies having low dust content and so fewer bins being masked in general, and non-axisymmetric structure giving an upper bound on how well-fit these galaxies can be. 

Future studies incorporating radiative transfer modelling or higher-resolution multi-band imaging could help disentangle the effects of dust and stellar populations on observed kinematics in regions where the dust is optically thin. Extending this analysis to the full GECKOS sample of edge-on galaxies, with a range of structural properties and inclinations, will provide stronger statistical constraints on when and where JAM residuals can be reliably used to detect kinematic structure. For example, the three galaxies in our sample with previously identified bars (NGC\,3957, IC\,1711, NGC\,0522) are \update{the three least dusty galaxies}, so a sample of galaxies with higher dust content as well as kinematic structure would provide new insights into the upper limit on dust extinction where kinematic components can still be identified. Additionally, a comparison of our work to simulations will allow a more confident determination of whether residual structure truly corresponds to non-axisymmetric orbits. Finally, this work provides a starting point for the GECKOS dynamical modelling effort, and the lessons learned on dust masking and MGE fitting will inform future papers. Future GECKOS studies will employ more complex techniques such as orbit-superposition modelling to explicitly model stellar bars \citep{2022ApJ...941..109T,2024MNRAS.534..861T} and derive parameters such as bar pattern speed. \update{Alternative approaches are also possible, such as the method of \cite{2005A&A...430...67F}, who derived bar pattern speeds from H$\alpha$ residual velocity maps with appropriate dust masking.}

We emphasise the importance of caution when applying axisymmetric models to edge-on galaxies. Dust along the line of sight affects the observed kinematics, especially in the mid-plane, tracing only a subset of stars. While using near-infrared imaging helps reduce this bias in the mass model, this imaging introduces a mismatch in our stellar tracer model when compared to optical kinematics, which appears, for example, in poor modelling of nuclear discs. However, the impact of this mismatch between tracer wavelengths could be improved in future by allowing M/L to vary with each luminous Gaussian, for example. Despite these limitations, global parameters like enclosed mass and inclination are robustly recovered, even under aggressive masking. \update{A mask of \update{E(B–V)} > 0.7 is sufficient for stable global results, whereas the stricter mask of \update{E(B–V)} > 0.2 should be applied when using residual maps for diagnosing non-axisymmetric structure.} This suggests that independent of their kinematic complexity, dusty star-forming edge-on discs can be reliably modelled axisymmetrically at a global level, while residual maps can serve as a window into their more complex kinematic substructure, shedding light on the internal dynamics that trace their evolution.

\begin{acknowledgements}
      This work is based on observations obtained with ESO telescopes at the La Silla Paranal Observatory under programme ID 110.24AS. We gratefully acknowledge the support of the ESO staff, and in particular the dedicated team at Paranal Observatory, for their efforts in executing the GECKOS observations.

      This research was partially supported by the Australian Research Council Centre of Excellence for All Sky Astrophysics in 3 Dimensions (ASTRO 3D), through project number CE170100013. THR acknowledges the support and funding of an ESO Studentship. AFM acknowledges the support and funding of an ESO Fellowship. AP acknowledges support from the Hintze Family Charity Foundation. MM acknowledges support from the UK Science and Technology Facilities Council through grant ST/Y002490/1. DAG acknowledges support from the UK Science and Technology Facilities Council through grant ST/X001075/1. FP acknowledges support from the Horizon Europe research and innovation programme under the Maria Skłodowska-Curie grant “TraNSLate” No 101108180, and from the Agencia Estatal de Investigación del Ministerio de Ciencia e Innovación (MCIN/AEI/10.13039/501100011033) under grant (PID2021-128131NB-I00) and the European Regional Development Fund (ERDF) ``A way of making Europe'’. PD is supported by a UKRI Future Leaders Fellowship (grant reference MR/S032223/1). 
\end{acknowledgements}

\bibliography{geckos-bib}{}
\bibliographystyle{aa}
\begin{appendix}

\section{JAM models}
\begin{figure}[h]
    \centering
    \onecolumn \includegraphics[width=\textwidth]{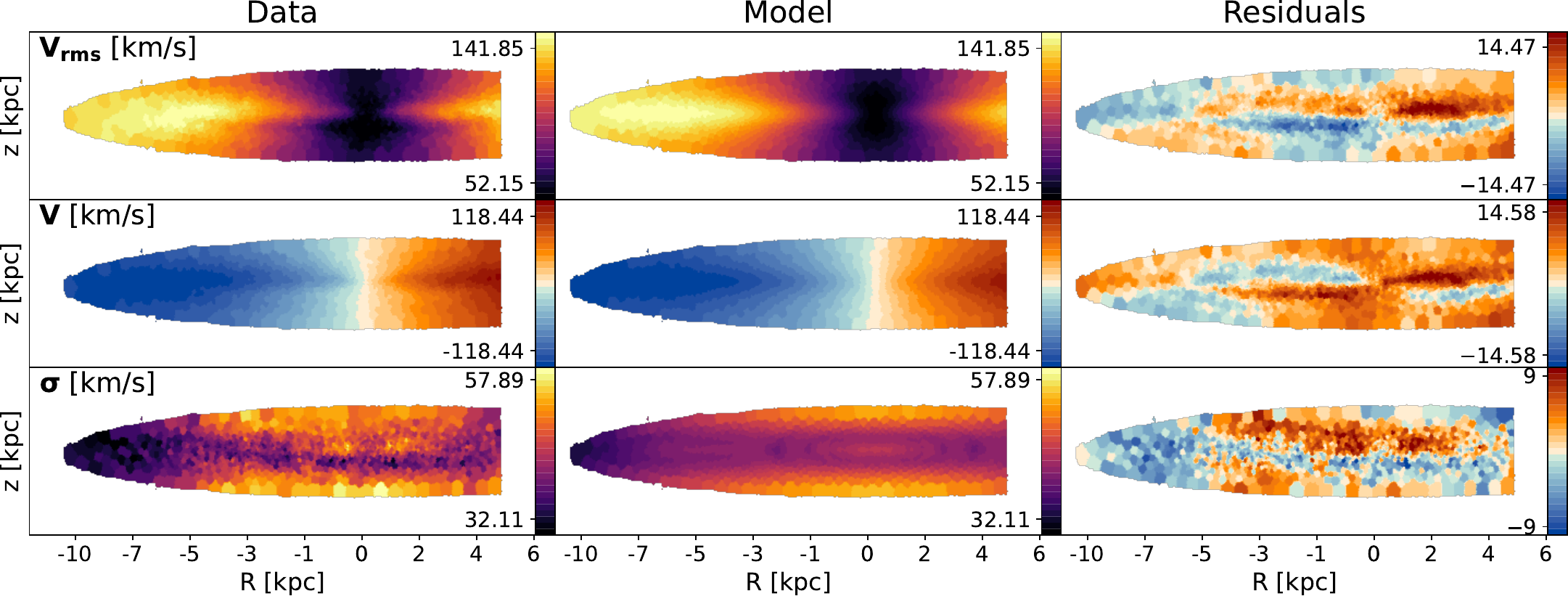}
    \caption{As for Figure \ref{fig:NGC3957}, but for NGC\,3279, and without circling structure.}
    \centering
    \label{fig:NGC3279}
\end{figure}

\begin{figure}[h]
    \centering
    \onecolumn \includegraphics[width=\textwidth]{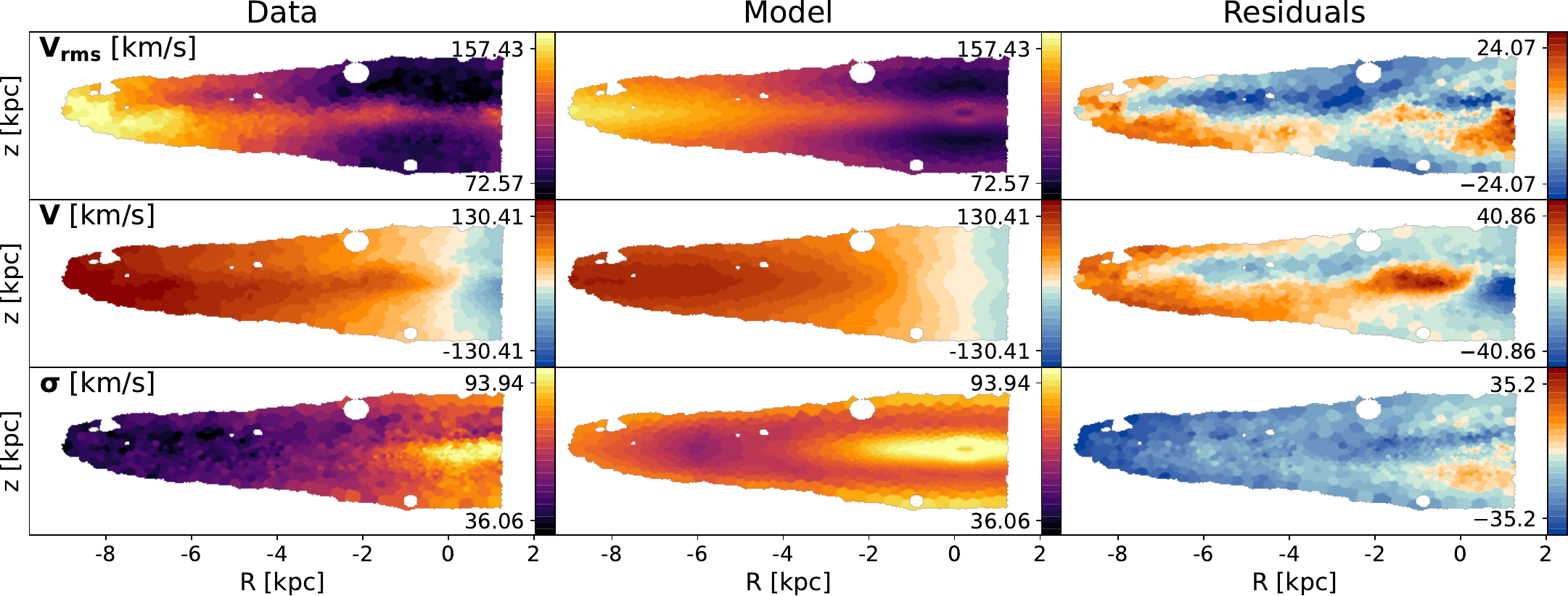}\caption{As for Figure \ref{fig:NGC3957}, but for NGC\,5775, and without circling structure.}
    \centering
    \label{fig:NGC35775}
\end{figure}

\begin{figure*}
    \centering
    \includegraphics[width=\linewidth]{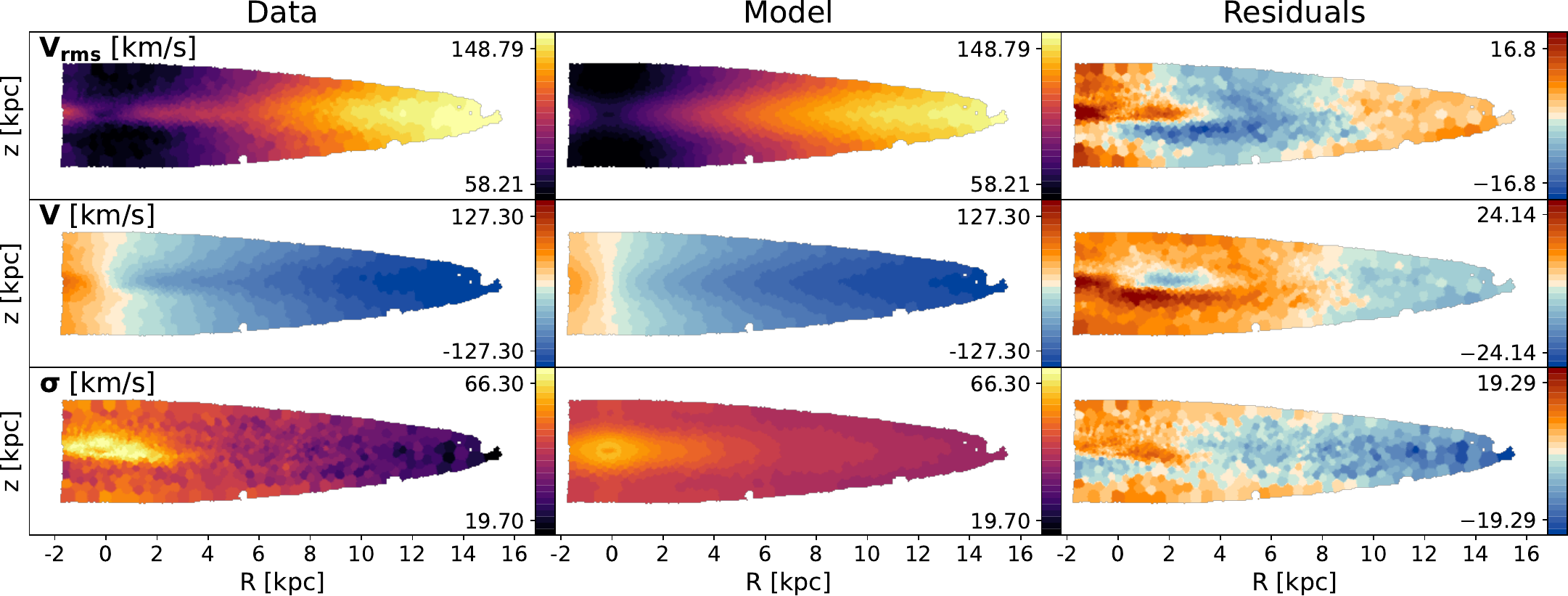}
    \caption{As for Figure \ref{fig:NGC3957}, but for NGC\,0360, and without circling structure.}
    \label{fig:NGC0360}
\end{figure*}

\begin{figure*}
    \centering
    \includegraphics[width=\linewidth]{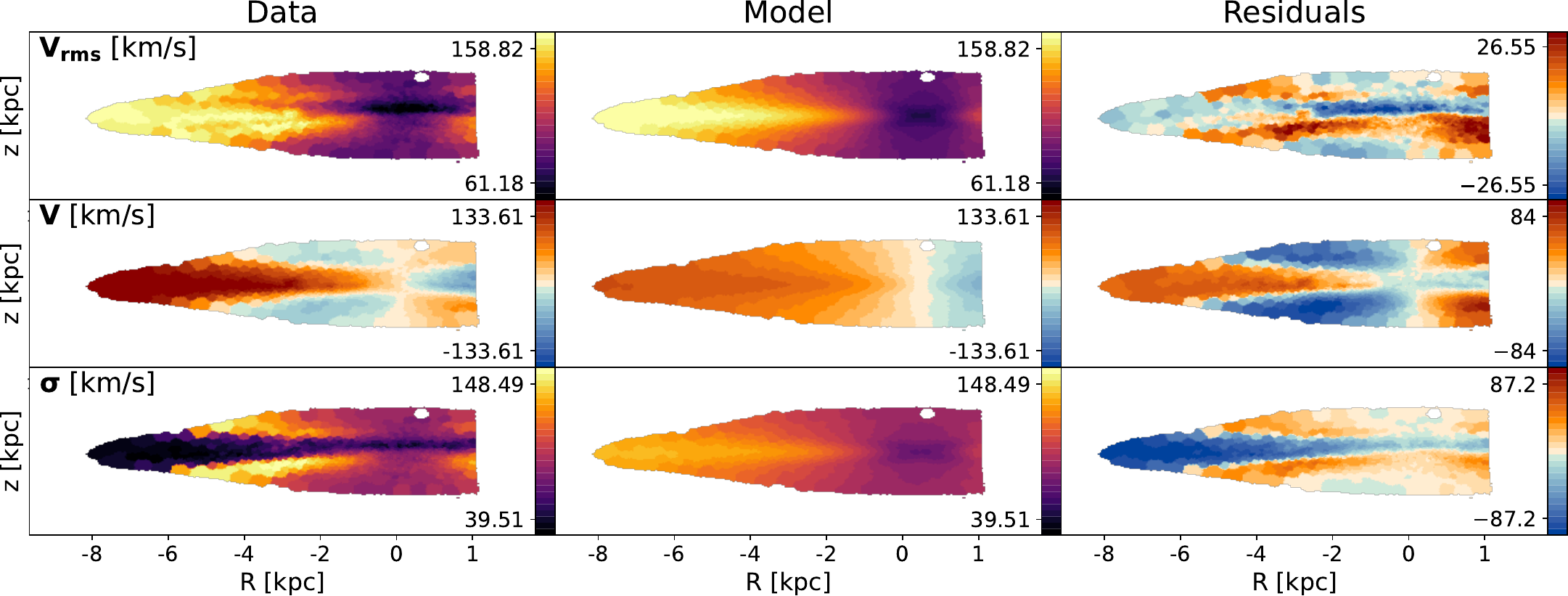}
    \caption{As for Figure \ref{fig:NGC3957}, but for UGC\,00903, and without circling structure.}
    \label{fig:UGC00903}
\end{figure*}

In this section, we present JAM model maps for galaxies NGC\,3279, NGC\,5775, NGC\,0360 and UGC\,00903, in Figures \ref{fig:NGC3279}-\ref{fig:UGC00903}. The left column shows $V_{\text{rms}}$, $V$ and $\sigma$ derived from \ngist\ output for the seven GECKOS galaxies, the central column shows the same but from the JAM model, and the right column shows the residuals, i.e. data minus model. The velocity residuals for these galaxies do not show clear non-axisymmetric structure.

\end{appendix}
\end{document}